\documentclass[10pt,letterpaper]{article}

\usepackage[T1]{fontenc}
\usepackage{XCharter}               
\usepackage[xcharter]{newtxmath}    
\usepackage[margin=1in]{geometry}
\usepackage{booktabs}
\usepackage{graphicx}
\usepackage{subcaption}
\usepackage{amsmath}   
\usepackage[numbers,sort&compress]{natbib}
\usepackage[colorlinks=true,linkcolor=blue,citecolor=blue,urlcolor=blue]{hyperref}
\usepackage{microtype}
\usepackage{xspace}
\usepackage{array}
\usepackage{pgfplots}
\pgfplotsset{compat=1.18}

\title{%
  Decomposing Runtime, Kernel, and Quantization Speedups\\
  via a Matched FP16 Intermediate:\\
  A Hardware-Conditioned Case Study on Four NVIDIA RTX A5000 GPUs%
}

\author{%
  Weijia Han \qquad Lisha Qu\\[3pt]
  University of Washington%
}

\date{}

\begin{document}

\maketitle


\begin{abstract}
Reported serving speedups from quantized kernels typically bundle the
weight format, the kernel, and the inference runtime into one number. We
present an attribution study on four NVIDIA RTX A5000 GPUs, 24 GiB each,
on a single host with NVLink-bridged pairs. A matched intermediate stack
that keeps the faster runtime without the quantized kernel splits the
full speedup into a runtime part and a kernel and quantization part.
Under matched greedy decoding the full stack reaches $2.58\times$ end to
end, with the runtime change accounting for about two thirds of that
gain on a logarithmic scale; across three similar model families the
kernel and quantization part moves by at most 1.5\%. Sharding one
instance across all four cards falls well below doubling: a profiler
trace attributes about 80\% of the per token shortfall to coordination,
and an NVLink versus PCIe control on the same hardware shows similar
realized bandwidth on both links, pointing away from link bandwidth as
the cause. Whether to run one sharded instance or several independent
ones depends on the workload and the model, with the ranking reversing
on the larger model: the smaller model splits between sharding and
multiple instances by workload, while the larger model favors two paired
instances on every workload. Quantization extends sustainable concurrent
users roughly four times past a reproducible half precision memory
cliff. Differences in sampling mode and prompt pool between the two
stacks are documented as threats to validity.
\end{abstract}


\begin{figure}[t]
\centering
\includegraphics[width=0.98\linewidth]{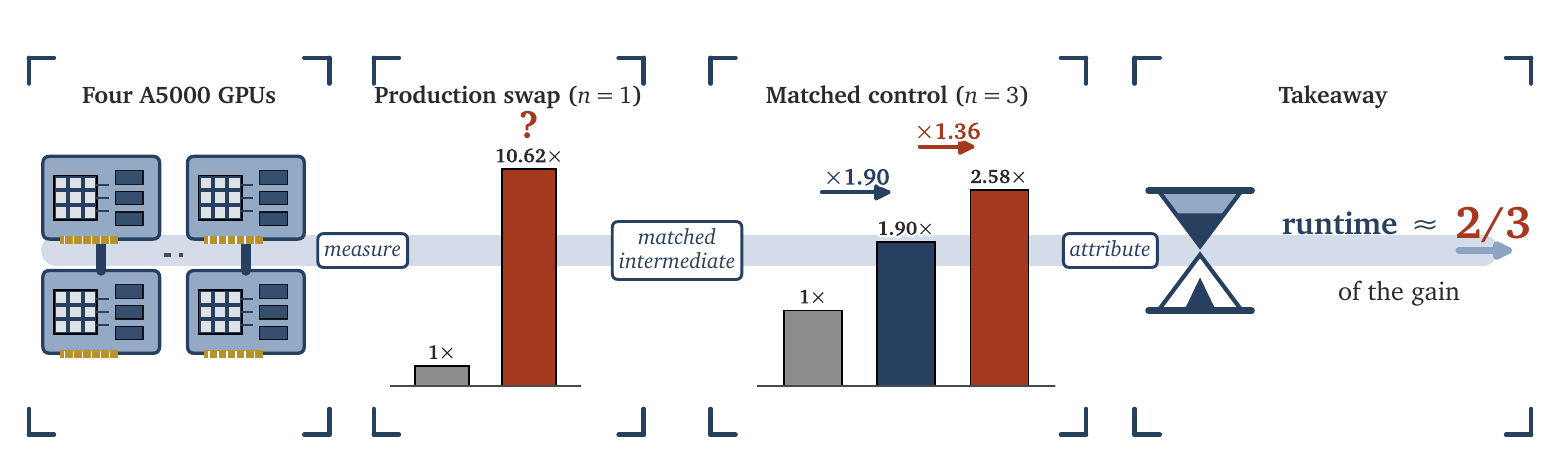}
\caption{Overview of the study. Four A5000 GPUs bridged into NVLink
pairs serve the same model through three stacks. A production swap
measured in a single run reports one conflated $10.62\times$ end to end
speedup, which bundles the runtime change, the kernel and quantization
change, and differences in batching and sampling. Inserting an
intermediate stack that keeps the faster runtime and drops the
quantized kernel, under matched greedy decoding with three runs, splits
the gain into a $1.90\times$ runtime factor and a $1.36\times$ kernel
and quantization factor, an overall $2.58\times$, with the runtime
carrying about two thirds of the gain on a logarithmic scale
(Table~\ref{tab:c1-sensitivity}).}
\label{fig:teaser}
\end{figure}


\section{Introduction}
\label{sec:introduction}

The deployment of large language models in production has motivated a
cascade of systems optimizations: weight quantization to reduce memory
footprint, paged attention to increase key value cache utilization,
continuous batching to keep accelerators saturated, and tensor parallel
sharding to spread computation across multiple accelerators. Each
technique is typically evaluated as a complete stack replacement. A
common framing is: ``switch from HF Transformers to vLLM-Marlin and
throughput grows ten times.'' This conflates weight format, GEMM
kernel, and inference runtime at once. These conflations make it hard
to answer practical questions: which part of the stack is doing the
work, and whether tensor parallel sharding still helps once a model
fits on a single accelerator.

We present an attribution study for the eight billion parameter Llama
model on four NVIDIA RTX A5000 cards, bridged into pairs and drawn from a
single host, a mid range configuration common in academic and
industrial clusters. By introducing a vLLM-FP16
intermediate, we obtain a multiplicative decomposition of the total
speedup into a \emph{runtime factor} and a \emph{kernel and
quantization factor}. Their product equals the overall speedup exactly
by construction; the matched intermediate cancels algebraically.
Figure~\ref{fig:teaser} previews the central result. We use
the decomposition as a bookkeeping device, and make no causal
independence claim (Section~\ref{subsec:decomposition}).

\begin{figure}[t]
\centering
\includegraphics[width=\linewidth]{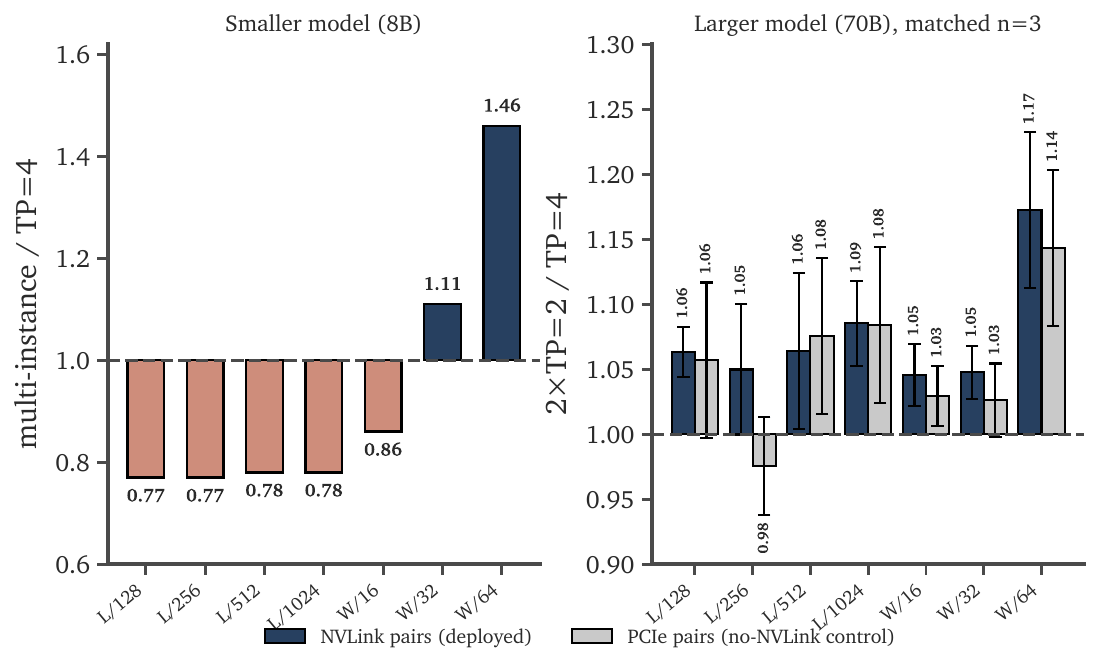}
\caption{Routing summary on the tested hardware. On the left, for the
smaller model, sharding one instance across all four cards wins the long
output workloads, shown by the bars below one, while running several
independent instances wins the wide workloads from the middle of the
range onward, shown by the bars above one. At the largest wide workload
the single sharded instance runs out of memory, leaving the independent
instances the only option. On the right, for the larger model, two
smaller instances are compared against one instance sharded across all
four cards, with the two instances placed either on the fast bridge, in
navy, or on the slower link, in gray. The two instances win on the
deployed hardware, and they keep essentially the same advantage when
moved to the slower link, so it does not depend on the connection
between the cards. Error bars show the spread across runs. The per
workload figures are in Table~\ref{tab:tp70b}.}
\label{fig:routing-summary}
\end{figure}

Our four main contributions are:

\begin{enumerate}
  \item \textbf{Attribution decomposition through a matched intermediate.}
        Matched greedy decoding yields a \textbf{2.58$\times$} end to
        end speedup at the wide batch operating point. The runtime swap
        accounts for roughly \textbf{two thirds} of the total log
        speedup under our log share metric (Eq.~\ref{eq:logshare}).
        Across the full concurrency sweep the runtime factor rises and
        the kernel and quantization factor shrinks
        (Table~\ref{tab:c1-envelope}). The kernel and quantization arm
        replicates across the tested GQA model family at the headline
        cell. The maximum pairwise spread is \textbf{1.5\%}
        (Table~\ref{tab:cross-model}, Section~\ref{subsec:cross-model}). As a deployment case study
        we also report the production stack swap; the apparent runtime
        factor inflates substantially under that setting and the gap
        reflects batching regime and sampling mode confounds, well beyond
        the runtime axis (Table~\ref{tab:c1-sensitivity},
        Section~\ref{subsec:controlled-rerun}).

  \item \textbf{Tensor parallel sharding does not stack on this
        hardware.} Sharding one instance across all four cards improves
        throughput well below the practical doubling threshold across
        every measured cell. A direct per output token latency
        measurement reports a material gap between the four card and
        single card arms at the wide batch operating point. A profiler
        trace on the same hardware attributes roughly \textbf{80\%} of
        that gap to the work of coordinating the shards at that point
        (Appendix~\ref{appendix:tp-napkin}). A control on the same
        hardware, pinning a pair of ranks first to the fast bridge and
        then to the slower bus, finds the two links deliver similar
        realized reduction bandwidth at our per step payloads, so the
        collective is bound by the cost of launching and synchronizing it
        at these sizes. The shortfall therefore points away from a
        bandwidth explanation; an arrangement that placed all four cards
        in a single fast switch fabric remains untested
        (Appendix~\ref{appendix:nvlink-projection}).

  \item \textbf{Multi instance routing is workload conditional and
        model conditional; sign reversal at the larger model.}
        Figure~\ref{fig:routing-summary} summarizes the routing outcome
        across all cells. For the smaller model there is a crossover:
        four way tensor parallel sharding wins long output cells, and
        multi instance routing wins wide batch cells from the medium
        wide batch cell onward. At the largest wide batch cell on the
        smaller model the four way sharded arm runs out of memory at
        engine initialization; multi instance routing remains the sole
        viable backend. For the larger model the outcome reverses on
        every workload. Two paired instances beat one sharded instance, a
        sign reversal versus the smaller model's long output cells. A
        profiler trace attributes the reversal to far fewer explicit
        reductions under the paired layout, since the smaller participant
        count moves most layer wise reductions onto a shared memory fast
        path (Appendix~\ref{appendix:tp-napkin}).

  \item \textbf{Capacity cliff outcome depends on runtime and kernel.}
        FP16 hits a memory cliff on the 24\,GiB cards in our cluster
        and the cliff reproduces in every independent run. Both INT4
        stacks extend the maximum sustainable concurrency by roughly
        \textbf{four times} without request failures. We report both
        standard per cell aggregate throughput and a bespoke deployment
        utility score defined and caveated in
        Section~\ref{subsec:capacity-protocol}. The two INT4 stacks
        differ substantially in per step throughput where both run, so
        the score ratio is reported as an operational gap between
        complete stacks, separate from the runtime axis.
\end{enumerate}

\paragraph{Scope, framing, and limitations.}
This paper focuses on attribution and routing implications
and does not attempt a full benchmark suite. All measurements come from a
single hardware configuration, four A5000 cards bridged into pairs, and
use the eight billion parameter Llama model with quantized weights, plus
the seventy billion parameter Llama model with quantized weights for the
larger model comparison, on a synthetic prompt pool. Sampling mode and
prompt pool asymmetries between the stacks are documented
(Section~\ref{subsec:sampling}). Detailed sample sizes and per cell
statistics appear in the body sections and appendices. A control on the
same hardware indicates the interconnect is unlikely to be the operative
bottleneck at our payloads (Appendix~\ref{appendix:nvlink-projection});
we do not claim generalization to a single fast switch fabric across all
four cards, to different GPU generations, to realistic workload
distributions, or to settings where greedy and ancestral decoding
produce materially different end to end timing. Section~\ref{sec:discussion} identifies
threats to validity and the conditions under which the tensor parallel
stacking verdict would reverse.

\paragraph{Paper organization.}
Section~\ref{sec:background} reviews the relevant background.
Section~\ref{sec:methodology} describes the experimental design,
including the documented asymmetries.
Section~\ref{sec:experiments} presents all four sets of results.
Section~\ref{sec:discussion} interprets the mechanistic causes and
limitations.
Section~\ref{sec:conclusion} provides hardware conditioned routing
implications.


\section{Background}
\label{sec:background}

\subsection{Autoregressive LLM Inference and Continuous Batching}

Autoregressive language model inference generates tokens one at a time
in a loop: each step runs a full model forward pass over the existing
key value (KV) cache and a single new input token (the \emph{decode} phase),
preceded by a potentially long \emph{prefill} phase that processes the input
prompt in one or a few batched passes.
The two phases have very different computational profiles: prefill is
compute bound (large matrix multiplications over the full prompt),
while decode is memory bandwidth bound (the model weights and the
growing KV cache must be read per token).

\emph{Continuous batching}~\cite{yu2022orca} improves GPU utilization
by scheduling iteration level batches: as soon as one sequence
finishes generation, a new one is admitted into the running batch
without waiting for all sequences to complete.
This eliminates the ``padding tax'' of static batching and keeps
GPU utilization high under heterogeneous request streams.

\subsection{PagedAttention and vLLM}

vLLM~\cite{kwon2023vllm} implements continuous batching combined with
PagedAttention, a KV cache management scheme inspired by operating
system virtual memory.
PagedAttention avoids contiguous per sequence buffers, which waste
memory on over provisioned maximum lengths, by dividing KV
memory into fixed size \emph{blocks} that are allocated on demand and
can be shared across sequences.
At wide concurrency, PagedAttention's block level allocation avoids
the contiguous padding waste that naive per sequence allocation incurs,
enabling significantly more in flight requests per GPU.

The vLLM asynchronous engine exposes this functionality through a
serving loop with a priority aware scheduler.
The engine we benchmark runs PagedAttention v2, which improves
multi head attention kernel efficiency over the original v1
implementation.

\subsection{Post-Training Quantization: GPTQ}

GPTQ~\cite{frantar2023gptq} is a one shot post-training quantization
method based on the approximate optimal brain surgeon (OBS) framework
that quantizes each weight row minimizing the layer wise reconstruction
error.
GPTQ reduces a sixteen bit model (FP16) to four bit integers (INT4) without
retraining, at the cost of a one time calibration pass.
For Llama-3.1-8B-Instruct, GPTQ-INT4 reduces weight memory from
approximately 16\,GiB to approximately 5.7\,GiB, a 2.8$\times$
reduction, while preserving generation quality on standard benchmarks.

A critical implementation detail is the choice of GEMM kernel at
inference time.
The naive approach uses a dequantize kernel that converts INT4 weights
to FP16 in registers on each call, adding substantial per call overhead
versus a direct FP16 matrix multiplication.
This is the root cause of a throughput regression we observed at an
earlier stage of this project: the same weights served through the
dequantize kernel ran slower than FP16.

\subsection{The Marlin INT4 GEMM Kernel}

Marlin~\cite{frantar2024marlin} is an FP16$\times$INT4 matmul kernel
designed for autoregressive LLM decode.
Marlin fuses the INT4 unpacking and dequantization into the GEMM
itself, avoiding a separate dequantization step before it, using a
specialized asynchronous pipeline that overlaps memory loads, scale
applications, and compute.
On NVIDIA Ampere GPUs, Marlin achieves near ideal four times speedup
over a standard FP16 matrix multiplication.

vLLM integrates the Marlin kernel natively via the Marlin INT4 backend,
making it directly accessible through the asynchronous engine without
changes to the model architecture or inference loop.

\subsection{Tensor Parallelism}

Tensor parallelism (TP)~\cite{shoeybi2019megatron} partitions weight
matrices across multiple GPUs along one dimension, splitting each
GEMM into per rank sub GEMMs followed by an allreduce collective
to aggregate results.
For a transformer layer, the column parallel projection and
row parallel projection can each be split across multiple ranks,
reducing per GPU peak memory proportionally and potentially
increasing throughput if the inter GPU communication cost is low
relative to the compute savings.

Two costs govern tensor parallel collectives: link bandwidth and the
fixed per collective launch and synchronization overhead.
On NVLink connected systems (e.g., H100 SXM, A100 SXM4),
allreduce bandwidth reaches approximately 600 to 900\,GB/s, so for
large payloads communication is nearly free.
On PCIe systems bidirectional bandwidth is lower (approximately
32\,GB/s on a sixteen lane PCIe gen 4 link), which can dominate the
allreduce cost for large payloads. For the small per token collectives
of INT4 decode, however, the fixed per collective overhead can dominate
instead, so a nominal bandwidth gap need not translate into a runtime
gap; we measure this directly on our hardware in
Appendix~\ref{appendix:nvlink-projection}.

\subsection{FlashAttention}
\label{subsec:bg-flash}

FlashAttention~\cite{dao2022flashattention} is an IO aware attention
kernel that tiles the attention computation to avoid materializing the
full quadratic attention matrix in GPU HBM, instead keeping
intermediate results in fast on chip SRAM.
vLLM's attention kernels are built on FlashAttention style
IO awareness principles; the specific kernel used in our experiments
is vLLM's PagedAttention v2 (a paged variant of the FlashAttention
computation).

\subsection{Grouped Query Attention}
\label{subsec:bg-gqa}

Llama-3.1-8B-Instruct uses grouped query attention
(GQA)~\cite{ainslie2023gqa}, which reduces the
number of KV heads relative to query heads by a factor of 4.
This has a significant practical consequence for memory modeling:
naive KV cache size estimates based on older Llama-2 figures
overestimate by four times.

\subsection{Related Work: Adjacent Serving Systems}
\label{subsec:related-work}

Several modern open source inference frameworks share design space
with vLLM and motivate the runtime versus kernel attribution we study.
\textbf{SGLang}~\cite{zheng2024sglang} combines a frontend
language for structured LLM programs with a backend runtime that
caches RadixAttention prefixes; in serving terms it is an
alternative continuous batching runtime to vLLM and would replace
the ``vLLM-Marlin'' arm of our decomposition.
\textbf{TensorRT-LLM}~\cite{nvidia_tensorrt_llm} is NVIDIA's
production targeted inference runtime; on NVLink connected
H100/A100 SXM hardware it is the strongest published baseline,
but its Marlin or AWQ equivalent INT4 kernels are GPU architecture
specific and we do not directly measure it on our
A5000s.
\textbf{Text Generation Inference (TGI)}~\cite{huggingface_tgi}
is HuggingFace's production deployed serving framework, widely
used in industry; like vLLM, it implements continuous batching
with PagedAttention, and would be a natural cross runtime control
in a future revision.

A related family of adjacent quantization systems reports similarly
large end to end speedups from implementations that bundle a
numerical method with a dedicated kernel and runtime together.
\textbf{LLM.int8()}~\cite{dettmers2022llmint8} uses mixed precision
decomposition to preserve a small set of outlier features in higher
precision while quantizing the remaining weights to eight bits.
\textbf{SmoothQuant}~\cite{xiao2023smoothquant} moves quantization
difficulty from activations into weights through a mathematically
equivalent transformation, so both operands admit efficient integer
matrix multiplication. \textbf{AWQ}~\cite{lin2024awq} protects a
small fraction of salient weight channels identified from activation
statistics, reaching accuracy close to GPTQ at comparable memory
savings. Each of these systems reports its headline number against
an unoptimized baseline, so the reported gain bundles the numerical
benefit of quantization together with the engineering benefit of a
fused kernel and an optimized runtime. The matched intermediate we
use to separate these effects for GPTQ and Marlin would apply
equally to any of these adjacent methods.

Two recent papers refine the scheduling axis we hold constant.
\textbf{Sarathi-Serve}~\cite{agrawal2024sarathiserve} eliminates
prefill and decode interference via chunked prefill and stall free
scheduling, addressing throughput latency tradeoffs that our
fixed schedule cells average over.
\textbf{DistServe}~\cite{zhong2024distserve} disaggregates prefill
and decode onto separate GPU pools, which would reframe the
four way sharded versus multi instance choice we measure as a phase aware
allocation problem.

A broader body of work addresses the scaling, partitioning, and
benchmarking questions our decomposition touches.
\textbf{Pope et al.}~\cite{pope2023efficiently} study scaling
transformer inference and the compute versus communication tradeoffs
of partitioning a model across accelerators, the same tradeoff our
tensor parallel measurements probe directly on PCIe hardware.
\textbf{AlpaServe}~\cite{li2023alpaserve} studies statistical
multiplexing with model parallelism, and
\textbf{Splitwise}~\cite{patel2024splitwise} studies phase splitting
between prefill and decode; both refine the sharding versus
replication space we measure at a smaller scale, on four cards
within a single host. \textbf{MLPerf Inference}~\cite{reddi2020mlperf}
is the standard benchmark methodology for the field; it reports
whole stack numbers, which motivates our call for intermediate
baselines in future serving benchmarks. The serving survey by
\textbf{Miao et al.}~\cite{miao2023towards} places our attribution
study within a broader taxonomy of serving optimizations.
Speculative decoding~\cite{leviathan2023fast} is an orthogonal
acceleration axis that we hold constant throughout.

To our knowledge, no prior peer reviewed work isolates the runtime versus
kernel split on hardware of the A5000 class through a matched FP16
intermediate. We view this as complementary to the SGLang, Sarathi-Serve,
and DistServe line: those works optimize \emph{within} a single inference
runtime, while we measure \emph{across} runtimes to expose how much of a
celebrated Marlin INT4 speedup is attributable to the runtime swap on this
hardware class.
What distinguishes our setup from prior cross runtime comparisons
(e.g., DistServe's vLLM baseline) is the matched FP16 intermediate
that holds the runtime axis fixed while varying the kernel. Without
this intermediate, runtime and kernel speedups are mathematically
inseparable from a single end to end ratio.


\section{Methodology}
\label{sec:methodology}

\subsection{Hardware and Software Stack}
\label{subsec:hardware}

All experiments run on a single physical server with eight NVIDIA RTX
A5000 cards, each holding 24 GiB of memory. The cards are
bridged into NVLink pairs, and the two cards within a pair reach each
other over NVLink while separate pairs reach each other over the slower
PCIe bus. The four cards used for any one tensor parallel configuration
therefore form two NVLink pairs, and communication between pairs crosses
PCIe, so a reduction that spans all four cards uses both links. A control on the same hardware
(Appendix~\ref{appendix:nvlink-projection}) finds that the realized NVLink and
PCIe bandwidths are close at our message sizes, so this arrangement does not
by itself explain the tensor parallel results. The host runs CUDA 12.4.

The software stack consists of two independently configured environments:

\begin{itemize}
  \item \textbf{HuggingFace baseline environment:} Python 3, the
        HuggingFace ecosystem, PyTorch 2.4. Inference runs through
        SwiftServe's custom batched decode scheduler (v0.5.6), which
        implements padded KV left fusion and continuous batching on
        top of the standard generation API. Execution is on a single
        A5000 card.

  \item \textbf{vLLM environment:} Python 3, vLLM 0.7.3, the
        HuggingFace ecosystem (pinned to avoid a tokenizer backend
        change introduced in a later minor release), PyTorch 2.4.
        Inference runs through vLLM's asynchronous engine and
        PagedAttention v2. Single card runs use the default single
        GPU mode. Tensor parallel experiments use four way sharding
        across all four A5000 cards.
\end{itemize}

Exact lockfiles for both environments ship in the artifact release;
the vLLM patch version is verifiable from the lockfile at submission time.

\subsection{Notation}
\label{subsec:notation}
Table~\ref{tab:notation} summarizes the symbols used in the
benchmark sweep and the attribution decomposition. Per cell artifact
paths and wrapper script names are catalogued in
Appendix~\ref{appendix:data-sources}.

\begin{table}[t]
\centering
\caption{Notation used throughout the paper.}
\label{tab:notation}
\small
\begin{tabular}{ll}
\toprule
Symbol & Meaning \\
\midrule
$c$ & Concurrency: number of in-flight requests per cell. \\
$T$ & Maximum output tokens per request. \\
$N$ & Number of independent runs aggregated for a cell. \\
TP=$p$ & Tensor parallelism with $p$ ranks (NCCL allreduce). \\
$N{\times}$TP=$p$ & $N$ independent TP=$p$ instances behind a router. \\
MI & Multi-instance: $N{\times}$TP=$1$ for 8B, $2{\times}$TP=$2$ for 70B. \\
\midrule
\multicolumn{2}{l}{\emph{Decomposition factors (Eqs.~\ref{eq:runtime-factor} through~\ref{eq:overall-factor}):}} \\
runtime factor       & vLLM-FP16 / HF-FP16 \\
kern+quant factor    & vLLM-Marlin / vLLM-FP16 \\
overall factor       & vLLM-Marlin / HF-FP16 \\
log-share$_i$        & $\ln(\text{factor}_i) / \ln(\text{overall factor})$ (Eq.~\ref{eq:logshare}) \\
\midrule
\multicolumn{2}{l}{\emph{Backend names:}} \\
HF-FP16        & HuggingFace transformers + SwiftServe scheduler, FP16. \\
HF-GPTQ        & HuggingFace transformers with the default quantized INT4 kernel. \\
vLLM-FP16      & The vLLM asynchronous engine, half precision weights. \\
vLLM-Marlin    & The vLLM asynchronous engine with the Marlin INT4 matrix multiply kernel. \\
\bottomrule
\end{tabular}
\end{table}

\subsection{Model}

We evaluate Llama-3.1-8B-Instruct~\cite{llama3herd2024} in two weight
formats:
\begin{itemize}
  \item \textbf{FP16:} The official Llama-3.1-8B-Instruct release
        (approximately 16\,GiB on GPU footprint).
  \item \textbf{GPTQ-INT4:}
        The upstream GPTQ-INT4 checkpoint of Llama-3.1-8B-Instruct
        (approximately 5.7\,GiB on GPU footprint, same architecture,
        GPTQ~\cite{frantar2023gptq} calibrated by the upstream
        publisher).
\end{itemize}
The GPTQ checkpoint disables descending activation order for compatibility
with the Marlin~\cite{frantar2024marlin} kernel path in vLLM.

\subsection{Prompt Pool}
\label{subsec:prompts}

All experiments use a fixed pool of \textbf{ten carefully curated short
prompts} covering technical Q\&A and creative writing tasks
(reproduced verbatim in Appendix~\ref{appendix:prompts}).
Mean tokenized length under the Llama-3.1 tokenizer is
\textbf{approximately 12 tokens}.\footnote{Length is approximate
because it depends on minor tokenizer configuration differences across the
SwiftServe and vLLM Python environments. The figure cited is from
the vLLM environment's default subword tokenizer.}
Prompts are cycled deterministically across all backends: the request
index selects one of the ten prompts in fixed order.
Both the SwiftServe baseline and the vLLM benchmark scripts use the
same ten prompt list (scripts committed to the artifact repository;
see appendix).

This design substantially reduces prompt distribution confounding by
holding the prompt list fixed across backends; we cannot rule out
residual prompt conditional effects at the tokenizer level.
However, a ten prompt synthetic pool is \emph{not} representative of a
realistic serving workload (e.g., ShareGPT, LMSys Chat 1M).
We list this as a threat to validity in
Section~\ref{sec:discussion}.

\subsection{Sampling-Mode Asymmetry}
\label{subsec:sampling}

The two backends were benchmarked with different decoding modes,
inherited from the upstream benchmark scripts:

\begin{itemize}
  \item \textbf{HF-FP16 served by SwiftServe} cells use
        \emph{deterministic greedy decoding}
        (greedy mode, temperature one). The benchmark script is
        committed to the artifact repository (see appendix).
  \item \textbf{vLLM} cells (FP16, Marlin, four way sharded Marlin)
        use \emph{ancestral sampling} (temperature one, top p equal
        to one, default vLLM sampling parameters). The benchmark
        scripts are committed to the artifact repository (see
        appendix).
\end{itemize}

This is a known asymmetry in our experimental design.
We argue that it does \emph{not} materially affect the throughput
conclusions for two reasons.
First, the per token cost of argmax greedy versus a categorical
multinomial draw (ancestral) is on the order of microseconds on
modern GPUs, whereas the per token GEMM and KV cache access cost is
on the order of milliseconds (our per decode step inter token
latency, computed as concurrent users divided by throughput from
Table~\ref{tab:tp-stacking}, is on the order of tens of milliseconds
at the wide batch operating point across four way sharded and single
GPU Marlin).
Sampling cost is therefore dominated by compute and cache; the choice
of sampling distribution has little influence.
Second, both modes generate exactly the maximum output tokens
per request because no early stopping criterion is active in
our benchmarks; total work and total throughput are determined by
the fixed token budget alone.

We acknowledge this is a pragmatic argument, weaker than a controlled
ablation.
A future revision should rerun all cells with both backends matched
on either mode (we recommend matching on greedy for determinism).
This is listed as a threat to validity in
Section~\ref{sec:discussion}.

\subsection{Benchmark Cells}

We sweep two workload shapes, totaling seven benchmark cells.
The \textbf{long output} shape covers four cells: concurrency 8 with
output token counts of 128, 256, 512, and 1024; the number of prompts
submitted is twice the concurrency level.
The \textbf{wide batch} shape covers three cells: 256 output tokens with
concurrency levels of 16, 32, and 64; the number of
prompts submitted is likewise twice the concurrency level.

\subsection{Throughput Definition}
\label{subsec:throughput-def}

Throughput is defined as
\begin{equation}
  \text{throughput (tok/s)} = \frac{\text{total output tokens}}{\text{wall clock seconds}}
\end{equation}
where the total output tokens counts all output tokens
across all completed requests in a cell, and the wall clock seconds measures
wall clock time from the first request submission to the last
completion.
Model load time is \emph{not} included; we report steady state inference throughput.

\subsection{Measurement Protocol}
\label{subsec:protocol}

\paragraph{Claim$\rightarrow$dataset map (for the reader).}
The paper reports numbers from three distinct measurement
regimes; this paragraph maps each numbered claim to the dataset
it relies on, so that the reader does not have to reconstruct
the mapping while reading.

\begin{itemize}
  \item \textbf{C1 operational (single run, mixed scheduler / sampling).}
        HF served via SwiftServe HTTP wrapper with
        the SwiftServe batched decode mode at a batch ceiling of eight
        and ancestral vLLM sampling; single run per cell.
        Headline: runtime $7.56\times$, K+Q $1.40\times$,
        overall $10.62\times$, $86\%$ log share at 64 concurrent users.
  \item \textbf{C1 controlled (three independent runs, matched greedy across seven cells and three models).}
        HF in process at batch size 64, all backends
        at temperature zero, three independent runs per
        backend per cell, replicated at the headline cell on
        Mistral-7B-v0.3 and Qwen2.5-7B.
        Headline: runtime $1.90\times$, K+Q $1.36\times$,
        overall $2.58\times$, $67.9\%$ log share at 64 concurrent users
        ($\ln 1.903 / \ln 2.581 = 0.6786$).
  \item \textbf{Post hardening multirun (five independent runs).}
        Same cell shapes as the controlled rerun, single
        backend at a time, used for the 8B four way sharded stacking
        sweep, the 8B multi instance comparison, and the 70B linear
        scaling diagnostic. Single GPU
        Marlin in this regime is $2{,}391.44$\,tok/s at 64 concurrent users;
        in the initial single-run campaign it is $2{,}559.18$\,tok/s. The
        two era discrepancy is documented in this section.
  \item \textbf{Capacity sweep (three independent runs at 96 or more concurrent users).}
        FP16 / HF-GPTQ / vLLM-Marlin at high concurrency;
        FP16 OOM cliff cross validated.
  \item \textbf{Follow-up measurements (mixed sample sizes).}
        Direct per token latency from the metrics exporter (three runs),
        a profiler trace comparing the four card sharded arm against a
        single card (one run), a communication benchmark on the same
        hardware comparing the fast bridge against the slower link (one
        run per message size), and the matched larger model comparison of
        one sharded instance against two paired instances (three runs per
        workload).
\end{itemize}

The operational (single run) and controlled (three independent runs) C1 regimes
both appear in the headline C1 reporting because they answer
different deployment questions; the post hardening multirun is
the load bearing data for C2 and C3.

The attribution decomposition from the initial single-run campaign uses a single run
per cell, and we acknowledge that single run measurements
are insufficient for statistical inference.
The 8B four way sharded stacking sweep was repeated at five independent runs under
post hardening benchmark hygiene.
The 8B four card single GPU multi instance comparison was
likewise re measured at five independent runs.
The 70B comparison of one sharded instance against two paired instances
is a matched control of three independent runs per workload, measured
together in a single session; a separate linear scaling diagnostic
reuses the earlier five run measurement of a single paired instance.
The FP16 capacity cliff sweep was repeated at three independent runs.

The attribution decomposition from the initial single-run campaign validates the
kernel and quantization factor against the post hardening five run
multirun. At 64 concurrent users the Marlin/FP16 ratio reads
$1.27\times$ (post hardening five independent runs),
$1.36\times$ (controlled matched greedy three independent runs,
Section~\ref{subsec:controlled-rerun}), and
$1.40\times$ (initial campaign, single run); a $1.27$ to $1.40$ range
($\sim 10\%$ spread). The extended controlled regime envelope
across the full seven cell sweep at three independent runs (Table~\ref{tab:c1-envelope})
shows that this $\sim 10\%$ same cell spread is one slice of a
broader concurrency driven trajectory: the kern+quant factor
shrinks monotonically from $\approx 2.4\times$ at concurrency 8
long output to $1.36\times$ at 64 concurrent users wide batch as the
scheduler / attention / KV management share of per step time
grows. We therefore describe the kern+quant factor as
``concurrency dependent within the controlled regime, and
directionally similar across regimes at fixed concurrency'';
the runtime arm log share remains the dominant term at high
wide batch concurrency under both the controlled and the
operational regimes (Table~\ref{tab:c1-envelope}, Tables~\ref{tab:attribution-full} and~\ref{tab:dominance}).
Cross model headline cell replication (three independent runs, matched greedy)
on three GQA style seven to eight billion parameter models (Llama-3.1-8B,
Mistral-7B-v0.3, Qwen2.5-7B) bounds the residual model axis
contribution to the kern+quant factor at $\leq 1.5\%$
(Section~\ref{subsec:cross-model}); model axis variance is
therefore not the source of the $\sim 10\%$ same cell spread
seen across regimes.

\paragraph{Note on two single card baselines at the widest batch.}
The reader will encounter two slightly different single card throughput
numbers for the optimized stack at the widest batch. One comes from the
attribution measurement and the other from the later stacking and routing
measurements, which were taken in a different benchmark era. We keep both
because each is matched to its own baseline within its era, and the
difference between the two eras is one part of the small spread in the
kernel and quantization factor noted above. The corresponding tables
identify which number each result uses.

We did not pin the card clocks, did not enforce idle time between
workloads, and did not control for heat building up over sustained
batches. Spot checks during development suggested negligible drift over a
short sweep on a quiet host with default cooling, but this is not a
controlled measurement and should not be read as a formal stability
bound.

\subsection{Multiplicative Decomposition Design}
\label{subsec:decomposition}

The core methodological contribution is the \emph{matched intermediate}:
by running vLLM with FP16 weights (no quantization, standard
unquantized matrix multiplication) in addition to vLLM-Marlin and
HF-FP16, we obtain three measurement points per cell.
We define three derived ratios:
\begin{align}
  \text{runtime factor}    &= \frac{\text{vLLM-FP16}}{\text{HF-FP16}}
    \label{eq:runtime-factor} \\
  \text{kern+quant factor} &= \frac{\text{vLLM-Marlin}}{\text{vLLM-FP16}}
    \label{eq:kern-factor} \\
  \text{overall factor}    &= \frac{\text{vLLM-Marlin}}{\text{HF-FP16}}
    \label{eq:overall-factor}
\end{align}

\paragraph{The product law is an identity, true by construction.}
By the definitions in
Eqs.~\ref{eq:runtime-factor} through~\ref{eq:overall-factor}, the runtime
factor and the kern+quant factor share a common term (vLLM-FP16) that
cancels exactly:
\begin{equation}
  \frac{\text{vLLM-Marlin}}{\text{vLLM-FP16}} \cdot
  \frac{\text{vLLM-FP16}}{\text{HF-FP16}} =
  \frac{\text{vLLM-Marlin}}{\text{HF-FP16}}.
  \label{eq:product-law}
\end{equation}
Hence \emph{by construction} the product equals the overall factor at
machine precision; we do not present this as evidence of factor
independence, and any small numerical discrepancy in the reported
table values is due to rounding of the displayed decimals.
The decomposition is therefore a definitional identity, useful as a
\emph{bookkeeping} device that splits one observed ratio into two
ratios with the matched intermediate vLLM-FP16; it does not validate
the assumption that the underlying runtime and kernel mechanisms act
on disjoint slices of wall time.
We label the runtime and kern+quant factors as descriptive names for
the two arms of the split, and make no claim that they are causally
independent contributions.

\paragraph{Factor share metric.}
\label{subsec:share-metric}
We report each factor's contribution using the \emph{log share}:
\begin{equation}
  \text{share}_i = \frac{\ln(\text{factor}_i)}{\ln(\text{overall factor})}.
  \label{eq:logshare}
\end{equation}
Log share composes correctly with multiplicative decomposition: if
$\text{overall} = \prod_i \text{factor}_i$ exactly, then
$\sum_i \text{share}_i = 1$.
A linear share (\,$\text{factor}_i / \text{overall}$\,) does not
compose this way and would over credit whichever factor is measured
first.
All ``X\%~share'' claims in this paper use
Equation~\ref{eq:logshare}.
The log share is a property of the chosen decomposition under the
identity in Eq.~\ref{eq:product-law}; it describes how the observed
overall factor splits into the two named ratios, and implies no
causal share of wall clock gain attributable to ``runtime'' or ``kernel''
mechanisms in isolation.

\paragraph{What the decomposition cannot attribute.}
The runtime factor conflates the continuous batching scheduler,
PagedAttention KV management, and optimized device kernels in vLLM
beyond the GEMM (for example, the PagedAttention v2 attention
kernel versus HuggingFace's standard attention path).
The kernel and quantization factor conflates the Marlin GEMM kernel
and the reduction in weight memory.
Fully decomposing these sub factors would require additional
intermediate measurement points (for example, HuggingFace running
Marlin INT4, or vLLM running Marlin on FP16), neither of which is
straightforward to construct in the current vLLM release.

\subsection{Controlled C1 Rerun: Matched Greedy at Three Independent Runs}
\label{subsec:controlled-rerun}

The original C1 decomposition (Section~\ref{subsec:attribution})
carries a first order threat to validity: the single run measurement
compared two \emph{different} experimental conditions simultaneously. HF-FP16 ran through
SwiftServe's HTTP wrapper with the SwiftServe batched decode mode (batch
ceiling of eight) and greedy decoding, while vLLM-FP16 and
vLLM-Marlin ran through the asynchronous engine with continuous batching at
64 concurrent users and ancestral sampling. The two confounds are:
(a)~\emph{batching strategy}: fixed ceiling of eight (HF/SwiftServe) versus
continuous batching up to 64 concurrent users (vLLM); and
(b)~\emph{sampling mode}: greedy (HF) versus ancestral (vLLM).

\paragraph{Controlled rerun design.}
We ran a matched greedy benchmark wrapper to fix both confounds
simultaneously (script path and commit hash in
Appendix~\ref{appendix:data-sources}).
Sampling parity: all three backends use deterministic argmax greedy
(temperature zero, top p equal to one for vLLM;
sampling disabled, 256 new tokens for HF); both are
deterministic token for token equivalents.
Backend parity caveat: HF is run in process at batch size 64, removing
SwiftServe's batch ceiling of eight; vLLM retains its
native asynchronous engine continuous batching at 64 concurrent users.
This matches the
effective batch size but does not fully equalize the scheduler architecture
(vLLM uses PagedAttention block allocation; HF uses padded KV fusion).

\paragraph{Sample size and precision.}
The rerun headline is 64 concurrent users at 256 output tokens,
128 prompts, three independent runs per backend.
All per backend CVs at the headline cell are well under the 5\% threshold
used throughout this paper.
The matched greedy rerun is additionally extended to the full
seven cell sweep at three independent runs per backend per cell (long output
at output token counts of 128, 256, 512, and 1024 with concurrency 8, and wide batch
at concurrency levels of 16, 32, and 64 with 256 output tokens).
The envelope is reported
as Table~\ref{tab:c1-envelope}; per cell CV across the seven
cells times three backends is $\leq 3.4\%$ for 20 of the 21
cells. One small cell host jitter outlier is not a measurement defect.
The headline cell C1 numbers are additionally re measured on two
non Llama model classes
(Section~\ref{subsec:cross-model}): Mistral-7B-v0.3 and
Qwen2.5-7B, each three independent runs matched greedy on the
asynchronous engine at the same headline cell.

\paragraph{Result summary.}
Per run measurements and an aggregated summary are committed to
the artifact repository (paths in
Appendix~\ref{appendix:data-sources}). Headline factors at the
wide batch operating point:
\begin{itemize}
  \item Controlled runtime factor: $\mathbf{1.90\times}$
  \item Controlled kernel+quant factor: $\mathbf{1.36\times}$
  \item Controlled overall factor: $\mathbf{2.58\times}$
  \item Controlled log share (runtime): $\mathbf{67.9\%}$
\end{itemize}

The roughly fourfold difference between the operational runtime factor
($7.56\times$) and the controlled runtime factor ($1.90\times$) is
consistent with removing the two confounds simultaneously.
We hypothesize that the gap is dominated by the batch ceiling removal
(changing from a ceiling of eight to batch size 64 shifts the GPU utilization regime
substantially). A smaller contribution likely comes from the sampling mode change;
an explicit factorial ablation (batch ceiling versus sampling) is left as
future work.
We do not treat one number as ``correct''; the operational figure
reflects what a practitioner switching from the SwiftServe HF stack to vLLM
would observe in production; the controlled figure reflects an isolated
runtime versus kernel attribution under matched conditions. Both are reported in
Table~\ref{tab:c1-sensitivity}.

\subsection{Four-Way Tensor Parallel Stacking Protocol}

For the four way sharded sweep, we fix the base configuration as vLLM-Marlin
(single GPU) and measure four way tensor parallel sharding across
all four A5000 GPUs for the same seven benchmark cells using the
same prompt pool.
The four way sharded run shards model weight matrices along the column dimension
on the projection inputs and the row dimension on the projection
outputs (standard Megatron style sharding~\cite{shoeybi2019megatron}).
The allreduce collectives use the standard collective communications
library~\cite{nvidia_nccl} across the pair bridged topology: NVLink
within each pair and PCIe gen 4 between the pairs.
We compute the TP stacking factor:
\begin{equation}
  \text{TP stacking factor} = \frac{\text{vLLM-Marlin-TP4}}{\text{vLLM-Marlin-1x}}
  \label{eq:tp-factor}
\end{equation}
and compare against the ideal $4\times$ and a practical $2\times$
threshold.

\subsection{Capacity Sweep Protocol}
\label{subsec:capacity-protocol}

For the capacity sweep, we measure three configurations at
concurrency levels above those covered by earlier phases.
HF-FP16 and HF-GPTQ are measured on the SwiftServe HF stack
(the default INT4 kernel, distinct from Marlin); for HF-GPTQ we
sweep concurrency levels of 96, 128, 192, and 256, while for HF-FP16 we measure
at 96 concurrent users
to establish the OOM cliff and do not attempt higher concurrencies.
The vLLM-Marlin capacity sweep additionally measures vLLM-Marlin (vLLM runtime plus Marlin
INT4 kernel) on the same four high concurrency cells at three independent runs to
support the conclusion's recommendation row.
The PyTorch expandable segment allocator
is active for the HF-stack runs to isolate true capacity exhaustion
from fragmentation effects; vLLM-Marlin uses the vLLM defaults
(no allocator override) and GPU memory utilization at 90 percent.
For the capacity comparison we report two quantities side by side:
(i)~steady state aggregate throughput (tok/s) at each feasible
concurrency, the standard serving throughput metric, and
(ii)~a \emph{bespoke deployment utility score} we call the
\emph{aggregate token throughput product}, defined as throughput
in tokens per second times concurrent users. The product score
deliberately multiplies an already aggregate throughput by the
concurrency level to combine throughput with concurrency reach in
a single number. We do not present this product as a
throughput improvement, since the result is a serving utility score, a different quantity from
a standard throughput measurement. We use it because the
C4 contribution is about extending the feasible concurrency range
past the FP16 memory cliff, and a comparison that holds concurrency
fixed cannot express that extension. The throughput only comparison
at fixed concurrency is
reported separately in Section~\ref{subsec:capacity}. Decomposing
the capacity regime gap into runtime and kernel components is not
feasible with the current data, because the comparison between
HF-GPTQ and vLLM-Marlin changes both axes simultaneously (HF-GPTQ
uses the default INT4 kernel, vLLM-Marlin uses
Marlin) and HF-FP16 cannot run at 96 or more concurrent users to provide a
controlled intermediate at high concurrency.


\section{Experiments}
\label{sec:experiments}

\subsection{Baseline: vLLM-Marlin vs.\ HF-FP16}
\label{subsec:phase12}

At the widest batch the fully optimized stack serves tokens about 10
times faster than the reference, which is what motivates splitting that
single number into a runtime part and a kernel and quantization part in
Section~\ref{subsec:attribution}.

Table~\ref{tab:phase12-baseline} reports the throughput of the two end
point configurations across all seven workloads. The reference is
HuggingFace transformers running the batched scheduler on one card, and
the optimized stack is vLLM with the quantized kernel on the same card.
Across the workloads the overall speedup ranges from about five times to
about 10 times, widening as the batch grows.

\begin{table}[t]
\centering
\caption{Baseline: HF-FP16 vs.\ vLLM-Marlin-INT4 throughput (tok/s).
All 7 cells: $n_{\mathrm{failed}}=0$.}
\label{tab:phase12-baseline}
\small
\begin{tabular}{llrrr}
\toprule
Shape & Cell & HF-FP16 & vLLM-Marlin & Overall \\
\midrule
Long-out & $c{=}8,\,T{=}128$  & 139.16 & 729.55  & $5.24\times$ \\
Long-out & $c{=}8,\,T{=}256$  & 144.55 & 740.22  & $5.12\times$ \\
Long-out & $c{=}8,\,T{=}512$  & 143.09 & 699.25  & $4.89\times$ \\
Long-out & $c{=}8,\,T{=}1024$ & 127.03 & 679.22  & $5.35\times$ \\
\midrule
Wide-batch & $c{=}16,\,T{=}256$ & 192.31 & 1349.57 & $7.02\times$ \\
Wide-batch & $c{=}32,\,T{=}256$ & 226.61 & 2123.28 & $9.37\times$ \\
Wide-batch & $c{=}64,\,T{=}256$ & 240.87 & 2559.18 & $10.62\times$ \\
\bottomrule
\end{tabular}
\end{table}

The overall speedup grows steadily as the batch widens, and it stays
roughly flat as the output length grows. This dependence on batch size is
what motivates the attribution analysis. A single dominant factor would
produce a speedup that does not change with batch size, so the growing
trend implies that one factor scales with the batch while the other does
not.

\subsection{Attribution Decomposition}
\label{subsec:attribution}

The matched intermediate split holds by construction. Under the
original operational measurement, taken from a single run, the
runtime part of the speedup strengthens steadily across the wide
batch sweep while the kernel and quantization part weakens, so
that at the widest batch the runtime change accounts for the
large majority of the overall speedup on the log scale. Under the
controlled rerun using matched greedy decoding and repeated three
times, the same headline cell shows a smaller but still runtime
dominated split: about two thirds of the speedup on the log scale
comes from the runtime part, with the remainder from the kernel
and quantization part. Both regimes are reported in
Table~\ref{tab:c1-sensitivity} and discussed below.

Table~\ref{tab:attribution-full} presents the full three way matrix
for all seven cells. The vLLM-FP16 intermediate throughput enables
the multiplicative decomposition described in
Section~\ref{subsec:decomposition}: the runtime factor equals
vLLM-FP16 throughput divided by HF-FP16 throughput
(Eq.~\ref{eq:runtime-factor}), and the kernel and quantization
factor equals vLLM-Marlin throughput divided by vLLM-FP16
throughput (Eq.~\ref{eq:kern-factor}).

\begin{table}[t]
\centering
\caption{Three way throughput matrix (tok/s) and decomposition factors.}
\label{tab:attribution-full}
\small
\begin{tabular}{llrrrrrr}
\toprule
Shape & Cell & HF-FP16 & vLLM-FP16 & vLLM-Marlin & Runtime & Kern+Quant & Overall \\
\midrule
Long & $T{=}128$  & 139.16 & 288.31 & 729.55  & $2.07\times$ & $2.53\times$ & $5.24\times$ \\
Long & $T{=}256$  & 144.55 & 302.49 & 740.22  & $2.09\times$ & $2.45\times$ & $5.12\times$ \\
Long & $T{=}512$  & 143.09 & 300.43 & 699.25  & $2.10\times$ & $2.33\times$ & $4.89\times$ \\
Long & $T{=}1024$ & 127.03 & 291.57 & 679.22  & $2.30\times$ & $2.33\times$ & $5.35\times$ \\
\midrule
Wide & $c{=}16$   & 192.31 & 574.54  & 1349.57 & $2.99\times$ & $2.35\times$ & $7.02\times$ \\
Wide & $c{=}32$   & 226.61 & 1042.91 & 2123.28 & $4.60\times$ & $2.04\times$ & $9.37\times$ \\
Wide & $c{=}64$   & 240.87 & 1822.06 & 2559.18 & $7.56\times$ & $1.40\times$ & $10.62\times$ \\
\bottomrule
\end{tabular}
\end{table}

\paragraph{Identity check.}
The runtime factor times the kernel and quantization factor equals the
overall factor exactly by construction (Eq.~\ref{eq:product-law}). Any
small discrepancy that appears when multiplying the two decimal values
shown in Table~\ref{tab:attribution-full} is a rounding artifact, and never
evidence that the factors are independent.

\paragraph{Headline finding.}
At the widest batch the runtime change is the larger of the two factors.
Applying the logarithmic share metric (Eq.~\ref{eq:logshare}) to the
factors at that operating point:
\begin{equation*}
  \text{share}_{\text{runtime}}(c{=}64)
    = \frac{\ln(7.56)}{\ln(10.62)}
    = \frac{2.023}{2.363}
    \approx 0.856,
\end{equation*}
which we report as about six sevenths of the speedup. This is the
operational figure; the controlled headline reported below is smaller.
Table~\ref{tab:dominance} summarizes how the runtime share grows as the
batch widens.

\begin{table}[t]
\centering
\caption{Runtime factor dominance under the log share metric
(Eq.~\ref{eq:logshare}).
At wide batch $c=64$, the runtime arm log share under this decomposition is 86\% of the
overall speedup on the log scale; at $c=8$ long output ($T=128$),
the two arms are nearly balanced.
Computed from Table~\ref{tab:attribution-full}.}
\label{tab:dominance}
\small
\begin{tabular}{lrrr}
\toprule
Cell & Runtime factor & Kern+Quant factor & Runtime log-share \\
\midrule
$c{=}8,\,T{=}128$ (long)   & $2.07\times$ & $2.53\times$ & $44\%$ \\
$c{=}8,\,T{=}1024$ (long)  & $2.30\times$ & $2.33\times$ & $50\%$ \\
$c{=}16$ (wide)            & $2.99\times$ & $2.35\times$ & $56\%$ \\
$c{=}32$ (wide)            & $4.60\times$ & $2.04\times$ & $68\%$ \\
$c{=}64$ (wide)            & $7.56\times$ & $1.40\times$ & $\mathbf{86\%}$ \\
\bottomrule
\end{tabular}
\end{table}

The runtime factor grows steadily as the batch widens, while the kernel
and quantization factor shrinks. Section~\ref{sec:discussion} offers
mechanistic hypotheses for both trends.

\paragraph{Operational and controlled attribution.}
Table~\ref{tab:c1-sensitivity} places the original single run production
numbers alongside the controlled rerun with matched greedy decoding
repeated three times (Section~\ref{subsec:controlled-rerun}).

\begin{table}[t]
\centering
\caption{Operational and controlled C1 attribution at $c{=}64$, $T{=}256$.
Operational: the production baseline served through the SwiftServe
wrapper at an eight request batch ceiling with ancestral sampling, one
run. Controlled: matched greedy decoding, baseline served in process at
batch size 64, three runs, coefficient of variation from about a
tenth of a percent to under two percent. The roughly fourfold difference
in the runtime factor is consistent with two confounds lifted at once:
raising the eight request batch ceiling, which we expect dominates, and
removing the sampling mode mismatch. An explicit factorial separation of
the two is future work. Both measurements are valid and answer different
deployment questions.}
\label{tab:c1-sensitivity}
\small
\begin{tabular}{lrr}
\toprule
Metric & Operational ($n{=}1$, mixed) & Controlled ($n{=}3$, matched greedy) \\
\midrule
HF-FP16 throughput (tok/s) & $240.87^{a}$ & $997.52 \pm 17.83$ \\
vLLM-FP16 throughput (tok/s) & $1{,}822.06^{b}$ & $1{,}897.96 \pm 2.44$ \\
vLLM-Marlin throughput (tok/s) & $2{,}559.18^{c}$ & $2{,}573.96 \pm 12.49$ \\
\midrule
runtime\_factor & $7.56\times$ & $\mathbf{1.90\times}$ \\
kernel+quant\_factor & $1.40\times$ & $\mathbf{1.36\times}$ \\
overall\_factor & $10.62\times$ & $\mathbf{2.58\times}$ \\
log\_share\_runtime & $86.0\%$ & $\mathbf{67.9\%}$ \\
\bottomrule
\end{tabular}
\smallskip

\noindent{\footnotesize
$^{a,b,c}$Operational sources: HF-FP16 served through the SwiftServe
wrapper with batched decoding enabled at an eight request ceiling;
vLLM-FP16 and vLLM-Marlin served through the asynchronous engine with
ancestral sampling. Artifact paths for all rows appear in
Appendix~\ref{appendix:data-sources}.}
\end{table}

\paragraph{Controlled envelope across the full sweep.}
We extended the controlled decomposition from the single headline
workload to all seven, with three runs for each stack on each workload.
The runs were stable, with low variation from run to run on all but one
workload. Table~\ref{tab:c1-envelope} reports the full controlled
trajectory, and its widest batch row matches the controlled column of
Table~\ref{tab:c1-sensitivity}.

\begin{table}[t]
\centering
\caption{Controlled C1 envelope, matched greedy $n{=}3$ across
$7$ cells. Throughputs are $n{=}3$ means in tok/s; factors are
derived from the per cell $n{=}3$ means. The wide batch
trajectory shows runtime$\,\uparrow$ and kernel+quant$\,\downarrow$
monotonically with concurrency, with the runtime log share
crossing $50\%$ between $c{=}32$ and $c{=}64$. The headline
$c{=}64$ row is the headline of this $7$-cell trajectory and is
the controlled column of Table~\ref{tab:c1-sensitivity}.}
\label{tab:c1-envelope}
\small
\begin{tabular}{llrrrrrrc}
\toprule
Shape & Cell & HF-FP16 & vLLM-FP16 & vLLM-Marlin & Runtime & K+Q & Overall & Rt log-share \\
\midrule
Long & $T{=}128$  & 249.35 & 307.19  & 741.56  & $1.23\times$ & $2.41\times$ & $2.97\times$ & $19\%$ \\
Long & $T{=}256$  & 247.59 & 308.56  & 739.86  & $1.25\times$ & $2.40\times$ & $2.99\times$ & $20\%$ \\
Long & $T{=}512$  & 227.10 & 306.27  & 723.14  & $1.35\times$ & $2.36\times$ & $3.18\times$ & $26\%$ \\
Long & $T{=}1024$ & 193.38 & 298.79  & 696.61  & $1.55\times$ & $2.33\times$ & $3.60\times$ & $34\%$ \\
\midrule
Wide & $c{=}16$   & 435.18 & 597.48  & 1{,}350.26 & $1.37\times$ & $2.26\times$ & $3.10\times$ & $28\%$ \\
Wide & $c{=}32$   & 706.89 & 1{,}115.49 & 2{,}047.52 & $1.58\times$ & $1.84\times$ & $2.90\times$ & $43\%$ \\
\textbf{Wide} & $\mathbf{c{=}64}$ & $\mathbf{997.52}$ & $\mathbf{1{,}897.96}$ & $\mathbf{2{,}573.96}$ & $\mathbf{1.90\times}$ & $\mathbf{1.36\times}$ & $\mathbf{2.58\times}$ & $\mathbf{68\%}$ \\
\bottomrule
\end{tabular}
\end{table}

The controlled trajectory has the same shape as the operational one in
Table~\ref{tab:attribution-full}: as the batch widens the runtime factor
rises, the kernel and quantization factor shrinks, and the runtime share
grows past half of the speedup at the widest batch. The controlled
trajectory is a shrunken version of the operational one along the runtime
axis, and the difference between them is the joint effect of the batch
ceiling on the reference stack and the difference in sampling between the
two stacks. The kernel and quantization factor tracks batch size
in a systematic way well above measurement noise, being larger on the
decode heavy workloads and smaller on the wide ones.

\subsection{Cross-Model Replication at the Headline Cell}
\label{subsec:cross-model}

At the widest batch the kernel and quantization part of the
decomposition barely moves across three architecturally distinct models
of roughly seven to eight billion parameters, changing by at most about
one and a half percent. This replicates the kernel and quantization part
alone and leaves the whole decomposition untested, because we do not run
the reference stack again on the other two models. Doing so would require porting
the batched scheduler to their architectures, and it is not needed to
test how stable that part of the decomposition is across models.

\paragraph{Setup.}
We measured the controlled comparison again at the widest batch on two
additional models, served by the same engine on the same machine, three
times each. One of the two models needed a small
configuration workaround for a known issue in the serving library, which
we describe in the code release. As above, we do not run the reference
stack on these models, because the purpose here is to test the stability
of the kernel and quantization part alone.

\paragraph{Result.}
Table~\ref{tab:cross-model} reports the throughput of each stack and the
kernel and quantization factor derived from them across the three models.

\begin{table}[t]
\centering
\caption{Cross-model headline-cell C1 replication at
$c{=}64,\,T{=}256$ (matched greedy, $n{=}3$, same $4{\times}$
A5000 PCIe gen4 hardware). The kernel+quant factor varies by
at most $1.46\%$ across the three model classes; the maximum
pairwise drift is between Llama-3.1-8B and Qwen2.5-7B.
Mistral-7B-Marlin CV $= 4.4\%$ is the largest of the six cells
and stays below the $5\%$ threshold used elsewhere in this
paper.}
\label{tab:cross-model}
\small
\begin{tabular}{lrrrr}
\toprule
Model & vLLM-FP16 (tok/s) & vLLM-Marlin (tok/s) & K+Q factor & drift vs.\ Llama \\
\midrule
Llama-3.1-8B-Instruct        & $1{,}897.96 \pm 2.44$  & $2{,}573.96 \pm 12.49$ & $\mathbf{1.356}$ & --- \\
Mistral-7B-Instruct-v0.3     & $2{,}018.51 \pm 52.89$ & $2{,}765.80 \pm 57.19$ & $\mathbf{1.370}$ & $+1.03\%$ \\
Qwen2.5-7B-Instruct          & $2{,}019.10 \pm 11.13$ & $2{,}778.12 \pm 64.15$ & $\mathbf{1.376}$ & $+1.46\%$ \\
\bottomrule
\end{tabular}
\smallskip

\noindent{\footnotesize K+Q factors are computed from the $n{=}3$
mean throughputs at full precision; drift is
$(\text{factor}_m - \text{factor}_{\text{Llama}}) /
\text{factor}_{\text{Llama}}$.}
\end{table}

The small drift is well inside any reasonable claim that the factor is
stable across models on this hardware. The two smaller models reach
nearly identical throughput on both stacks and both run a little faster
than the eight billion parameter model, consistent with their smaller
weights taking less time per token. This is an observation about absolute
throughput; the kernel and quantization ratio absorbs the proportional
gain because the quantized kernel's advantage tracks the weight size. We
treat this as evidence that the kernel and quantization part of the
decomposition is stable across similar models of this size on this
hardware. We do not rerun the reference stack on the other two models, so
we make no claim about the runtime part or the full decomposition beyond
the model measured in full, and we do not test much larger or very
different models.

\subsection{Tensor Parallel Stacking Across Four Cards}
\label{subsec:tp}

Sharding one instance across four cards raises throughput to only about
1.5 times that of a single card, well short of a practical doubling
threshold. Several independent instances win the
wide workloads, lose the long output workloads on the smaller model, and
win throughout on the larger model.

\begin{figure}[t]
\centering
\includegraphics[width=0.9\linewidth]{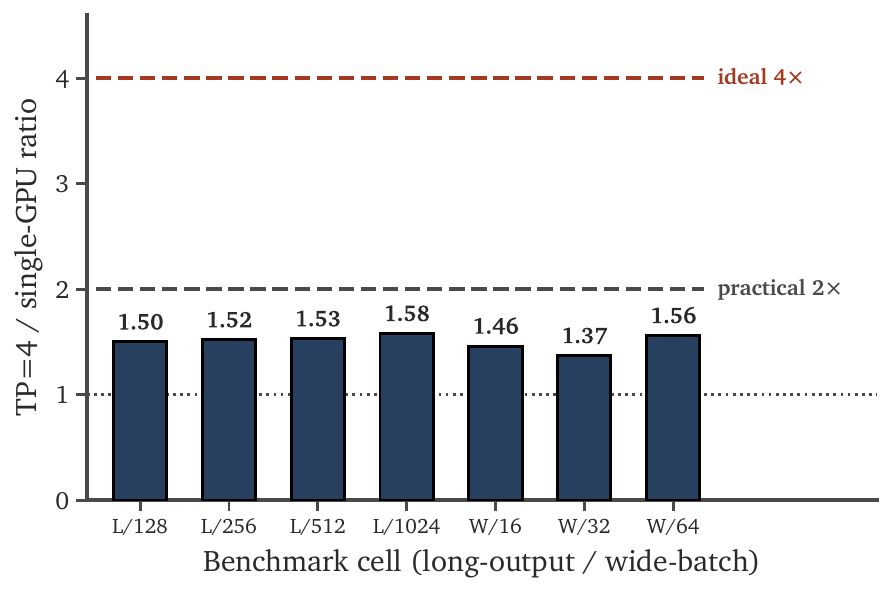}
\caption{Four way tensor parallel stacking factor across the
concurrency and output length sweep on the tested pair bridged topology.
No cell reaches the practical
doubling threshold (gray dashed); all fall far below the ideal four
times (terracotta dashed). Detailed values are in
Table~\ref{tab:tp-stacking}. The dip at the medium wide batch cell
relative to the large wide batch cell is discussed in
Section~\ref{subsec:why-tp-fails}.}
\label{fig:tp-stacking}
\end{figure}

Figure~\ref{fig:tp-stacking} plots, and Table~\ref{tab:tp-stacking}
reports, the stacking factor
(Eq.~\ref{eq:tp-factor}) for all seven workloads alongside the ideal of
four times.

\begin{table}[t]
\centering
\caption{TP=4 stacking on top of single GPU Marlin.
Post hardening $n{=}5$ multirun: TP=4 adds only $1.37\times$ to $1.58\times$
over single GPU; the ideal would be $4.00\times$.}
\label{tab:tp-stacking}
\small
\begin{tabular}{llrrrr}
\toprule
Shape & Cell & Marlin 1x & Marlin TP=4 & TP factor & Ideal \\
\midrule
Long-out & $T{=}128$  & 704.35  & 1{,}056.24 & $1.50\times$ & $4.00\times$ \\
Long-out & $T{=}256$  & 706.47  & 1{,}076.14 & $1.52\times$ & $4.00\times$ \\
Long-out & $T{=}512$  & 697.46  & 1{,}066.36 & $1.53\times$ & $4.00\times$ \\
Long-out & $T{=}1024$ & 654.08  & 1{,}030.42 & $1.58\times$ & $4.00\times$ \\
\midrule
Wide-batch & $c{=}16$ & 1{,}286.14 & 1{,}872.74 & $1.46\times$ & $4.00\times$ \\
Wide-batch & $c{=}32$ & 2{,}005.79 & 2{,}753.28 & $1.37\times$ & $4.00\times$ \\
Wide-batch & $c{=}64$ & 2{,}391.44 & 3{,}735.41 & $1.56\times$ & $4.00\times$ \\
\bottomrule
\end{tabular}
\end{table}

No workload reaches even a doubling: the geometric mean across the
seven cells is $1.49\times$ the throughput of a single card, with each
cell averaged over five runs, far below both
the ideal of four times and a practical doubling. The long
output workloads stack a little better than the wide ones on average, and
the smallest gain appears at one of the middle wide workloads
(Section~\ref{sec:discussion}).

\paragraph{Several independent instances against one sharded instance.}
We bring up four independent instances, one per card, behind a router
that sends each request to the least loaded instance.
Table~\ref{tab:multiinstance} reports the measured aggregate against the
single sharded instance at the same four card cost. The artifact paths
and the driver are catalogued in Appendix~\ref{appendix:data-sources}.

\begin{table}[t]
\centering
\caption{Measured multi instance ($4{\times}$ vLLM-Marlin host
processes, in process least busy routing) vs.\ TP=4 vs.\ single GPU
Marlin at equal four GPU cost. All three columns are $n{=}5$ from the
same repeated measurement campaign; MI per cell CV
$\leq 2.94\%$, TP=4 CV $\leq 4.93\%$, single-Marlin CV $\leq 4.06\%$.
The ``efficiency'' column is the ratio of multi instance aggregate
to $4{\times}$ single-Marlin throughput; $100\%$ denotes perfect
linear scaling.
$c{=}128$ is the only configuration where TP=4 runs out of memory at engine
initialization under the GPU memory budget; multi instance with $c{=}32$ per
replica sustains it.}
\label{tab:multiinstance}
\small
\begin{tabular}{lrrrrr}
\toprule
Cell & 1$\times$ Marlin & TP=4 & 4$\times$ multi-inst.\ & multi/TP4 & efficiency \\
\midrule
$c{=}8,\,T{=}128$  &  704.35 & 1{,}056.24 &  811.62 & $0.77\times$ & $28.8\%$ \\
$c{=}8,\,T{=}256$  &  706.47 & 1{,}076.14 &  830.04 & $0.77\times$ & $29.4\%$ \\
$c{=}8,\,T{=}512$  &  697.46 & 1{,}066.36 &  829.14 & $0.78\times$ & $29.7\%$ \\
$c{=}8,\,T{=}1024$ &  654.08 & 1{,}030.42 &  803.80 & $0.78\times$ & $30.7\%$ \\
$c{=}16,\,T{=}256$ & 1{,}286.14 & 1{,}872.74 & 1{,}612.04 & $0.86\times$ & $31.3\%$ \\
$c{=}32,\,T{=}256$ & 2{,}005.79 & 2{,}753.28 & 3{,}049.49 & $1.11\times$ & $38.0\%$ \\
\textbf{$c{=}64,\,T{=}256$} & \textbf{2{,}391.44} & \textbf{3{,}735.41} & \textbf{5{,}442.78} & $\mathbf{1.46\times}$ & $\mathbf{56.9\%}$ \\
$c{=}128,\,T{=}256$ &       --- & \emph{OOM} & 7{,}243.57 & MI only &       --- \\
\bottomrule
\end{tabular}
\end{table}

At the widest batch covered by all three configurations, four
independent instances serve about 1.5 times as many tokens as a single
sharded instance. Assuming the four instances simply add up would
overstate their throughput by more than half, and at lighter loads the
shortfall is larger still. The pattern fits contention for the shared
host, its processor, and its memory bus dominating the time whenever each
instance is only lightly loaded.

The choice of layout flips with the workload. On all four decode heavy
long output workloads, and at the narrowest wide workload, the single
sharded instance is the faster choice, the opposite of what simple
scaling would predict. The two layouts cross over in the middle of the
wide range. At the largest wide workload the choice collapses to the
independent instances, because the single shard runs out of memory at
startup while the instances continue to serve.

\paragraph{Relation to a prior prefill measurement.}
The stacking factors in Table~\ref{tab:tp-stacking} show that whatever
advantage sharding brings during the initial prompt processing does not
carry through to full inference on our decode heavy workload, which uses
a short prompt and a long output.\footnote{An internal report measured a
larger sharding advantage during prompt processing alone
(\cite{phase8_internal}); we cite it for context only and do not rely on
it for any claim here.} We discuss the mechanism in
Section~\ref{sec:discussion}.

\paragraph{Direct per token latency measurement.}
The latency inferred from throughput above folds in the one time cost of
processing the prompt. To isolate the steady state cost of each output
token, we measured the widest batch again and read the serving library's
own per token timing directly, repeating each layout three times. Spread
across four cards, each decode step finishes about seven milliseconds
faster than on a single card, which is the gain from adding the three
extra cards, though it is far short of the fourfold gain those cards
could in principle give. This direct measurement agrees with the one
inferred from throughput to a fraction of a millisecond, and the
mechanism discussion in Section~\ref{subsec:why-tp-fails} anchors on this
measured gap.

\subsection{Two Smaller Instances Outperform One Larger Shard at 70B}
\label{subsec:tp70b}

The larger model reverses the verdict from the smaller one. When a
model is too large to fit on a single card, splitting the four GPUs into
two smaller instances outperforms sharding one instance across all four,
and it does so on every workload we measured. The improvement is modest
but consistent, close to seven percent on average. The advantage holds, on
average, when the two instances use only the slower link, which shows the
interconnect explains little of it.

The larger model with quantized weights does not fit on a single 24
gigabyte card but fits on two. At equal four GPU cost we therefore
compare a single instance sharded across all four cards against two
independent instances that each span two cards and share one request
router. To keep the comparison fair, we measured all three
configurations in a single session on one idle group of four GPUs, using
the same server and the same sampling settings, repeating every workload
three times and reporting the median. This matched comparison replaces
an earlier figure that had divided the throughput of the paired
instances by a single instance baseline recorded in a different session.
That older baseline happened to run several percent slower on the wide
workloads and so overstated the advantage, reporting a gain of about
15 percent where the fair comparison shows about seven.

\paragraph{Measured throughput comparison.}
Table~\ref{tab:tp70b} reports the throughput of each configuration on
all seven workloads.

\begin{table}[t]
\centering
\caption{Throughput on the larger model, in tokens per second. All three
configurations were measured together in one session on one idle group
of four GPUs, using the same server and the same sampling, with three
runs per workload and the median shown. The paired layout is measured
twice: once with each instance on the fast bridge, and once, as a
control, with each instance on the slower link. Ratios are relative to
the single sharded instance.}
\label{tab:tp70b}
\small
\begin{tabular}{llrrrrr}
\toprule
Shape & Workload & One shard & \multicolumn{2}{c}{Two instances} & \multicolumn{2}{c}{ratio to one shard} \\
\cmidrule(lr){4-5}\cmidrule(lr){6-7}
      &          &           & NVLink & PCIe & NVLink & PCIe \\
\midrule
Long output & $c{=}8,\,T{=}128$  & 159.1 & 169.2 & 168.1 & $1.06$ & $1.06$ \\
Long output & $c{=}8,\,T{=}256$  & 166.2 & 174.5 & 162.2 & $1.05$ & $0.98$ \\
Long output & $c{=}8,\,T{=}512$  & 147.9 & 157.4 & 159.1 & $1.06$ & $1.08$ \\
Long output & $c{=}8,\,T{=}1024$ & 122.9 & 133.4 & 133.2 & $1.09$ & $1.08$ \\
\midrule
Wide batch & $c{=}16,\,T{=}256$ & 321.2 & 335.7 & 330.7 & $1.05$ & $1.03$ \\
Wide batch & $c{=}32,\,T{=}256$ & 619.1 & 648.5 & 635.3 & $1.05$ & $1.03$ \\
Wide batch & $c{=}64,\,T{=}256$ & 899.3 & 1{,}054.3 & 1{,}028.2 & $\mathbf{1.17}$ & $\mathbf{1.14}$ \\
\bottomrule
\end{tabular}
\end{table}

On the deployed hardware the paired layout wins every workload,
improving throughput by about seven percent on average, with the largest
gains at the widest batch. Holding the same two instances to the
slower link leaves the average advantage
essentially unchanged, at about five percent. On the slower link the
paired layout still comes out ahead on six of the seven workloads and
trails on only one, by a small margin. The advantage therefore comes
from spreading the work across fewer communicating cards; the link
between them explains little. Both the deployed configuration and the slower
link control appear in Table~\ref{tab:tp70b}, and
Appendix~\ref{appendix:nvlink-projection} gives the full interconnect
comparison.

\begin{figure}[t]
\centering
\includegraphics[width=\linewidth]{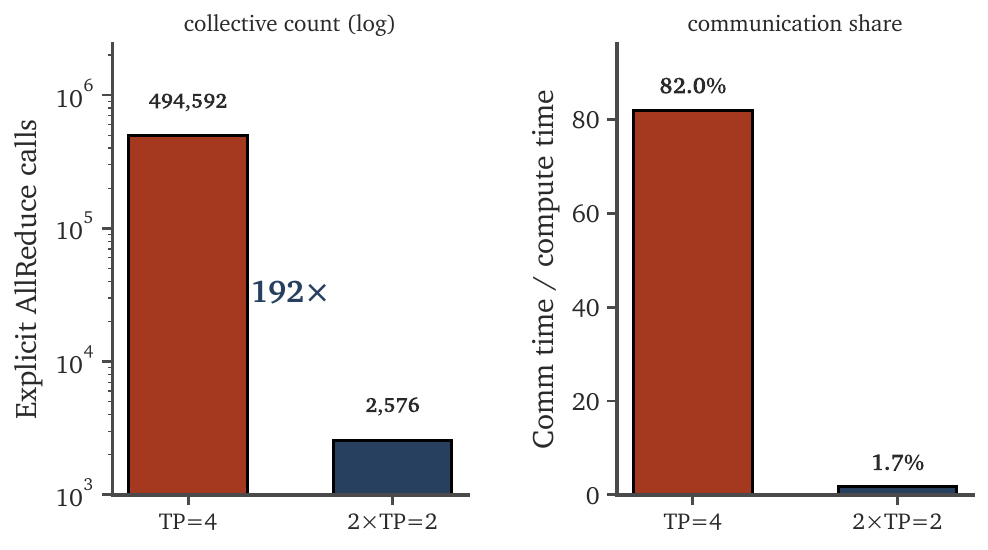}
\caption{Coordination collapses when fewer cards take part, shown for the
larger model at the widest batch. On the left, the number of explicit
reduction calls on a logarithmic scale: the single instance sharded
across all four cards issues on the order of two hundred times as many as
the two paired instances combined. On the right, the time spent
communicating as a share of the time spent computing, which is nearly as
large as computing for the single shard and almost none for the paired
instances. With fewer cards
per reduction the paired instances move the great majority of their
coordination onto a fast in memory path, which is why the larger model
favors several small instances over one large shard
(Section~\ref{subsec:tp70b}). The exact counts are in
Table~\ref{tab:tp-napkin-70b}.}
\label{fig:collective-collapse}
\end{figure}

This reverses what we saw on the smaller model, where a single instance
sharded across all four cards had won the long output workloads. The
cause is a hardware conditioned one. On the larger model the quantized
matrix multiplies over the bigger weights take longer per token, so the
fixed cost of coordinating the shards weighs less. Splitting into two
instances also halves the number of cards that must take part in each
reduction. A profiler trace at the widest batch makes the effect
concrete, as Figure~\ref{fig:collective-collapse} shows: the single
shard issues on the order of 200 times as many explicit
reduction calls as the two paired instances combined, and it spends
almost as much time coordinating as computing, while the paired
instances spend almost none. With only two cards per reduction the paired
instances also qualify for a fast in memory path that the single shard
cannot use. We develop this mechanism in Section~\ref{subsec:70b-reversal}.

\paragraph{Variance.}
The single shard is also the noisier configuration. Its throughput
varies more from run to run than the paired instances do, which fits the
picture that a larger set of communicating cards amplifies host side
timing jitter that smaller groups absorb. One first run of the single
shard at the widest batch was anomalously slow and is set aside by the
median. An earlier repeated measurement of the same comparison showed
the same pattern, and its run by run breakdown appears in
Appendix~\ref{appendix:70b-variance}.

\paragraph{Linear scaling gap.}
The next comparison is a separate diagnostic and does not bear on the
matched result above. It asks how far the paired throughput falls short
of simply doubling the throughput of one instance run alone, using the
earlier repeated measurement. Table~\ref{tab:tp70b-gap} reports the
shortfall.

\begin{table}[t]
\centering
\caption{70B Marlin INT4: measured $2{\times}$TP=2 throughput as a
fraction of naive linear extrapolation ($2\times$ single TP=2
$n{=}5$ mean throughput). Values below $1.0$ indicate overstatement
by linear scaling; the sole value above $1.0$ (wide\_c64) reflects
super linear gains from per instance KV pool saturation relief.}
\label{tab:tp70b-gap}
\small
\begin{tabular}{llr}
\toprule
Shape & Cell & meas / linear-est \\
\midrule
Long-out & $c{=}8,\,T{=}128$  & 0.50 \\
Long-out & $c{=}8,\,T{=}256$  & 0.49 \\
Long-out & $c{=}8,\,T{=}512$  & 0.49 \\
Long-out & $c{=}8,\,T{=}1024$ & 0.51 \\
\midrule
Wide-batch & $c{=}16,\,T{=}256$ & 0.53 \\
Wide-batch & $c{=}32,\,T{=}256$ & 0.64 \\
Wide-batch & $c{=}64,\,T{=}256$ & \textbf{1.11} \\
\bottomrule
\end{tabular}
\end{table}

The decode heavy workloads fall roughly a factor of two short of simple
doubling. When each instance handles only a few requests, its cache is
under used and the per token overhead on the host dominates relative to
the actual computation. The wide workloads fall short by less. The widest
is the sole exception, where the two instances together slightly exceed
twice one instance, because a single instance there is near the point
where its cache saturates and its throughput stops rising with load;
splitting the load places both instances in the more efficient part of
that curve.

The larger model falls far less short of doubling than the smaller model
does at the widest batch, because the heavier per token computation of
the larger model leaves contention on the host a smaller share of the
total time.

\subsection{Capacity Cliff and Aggregate Throughput}
\label{subsec:capacity}

The reference stack runs out of memory once the load passes 64
concurrent users, in every run. The capacity story then has two parts.
First, quantization roughly quadruples the number of concurrent users a
card can sustain before it runs out of memory, and both quantized stacks
reach the top of our range without a single failure. Second, on a
deployment utility score that rewards serving more users at once, both
quantized stacks far exceed the reference. For the first of them the
entire gain comes from serving more users, since at a fixed load it
actually trails the reference; the other is
faster at a fixed load as well and continues past the point where the
reference fails. The score gap between the two quantized stacks combines
a runtime change and a kernel change, and we report it as an operational
gap across the whole stack.

\begin{figure}[t]
\centering
\includegraphics[width=0.85\linewidth]{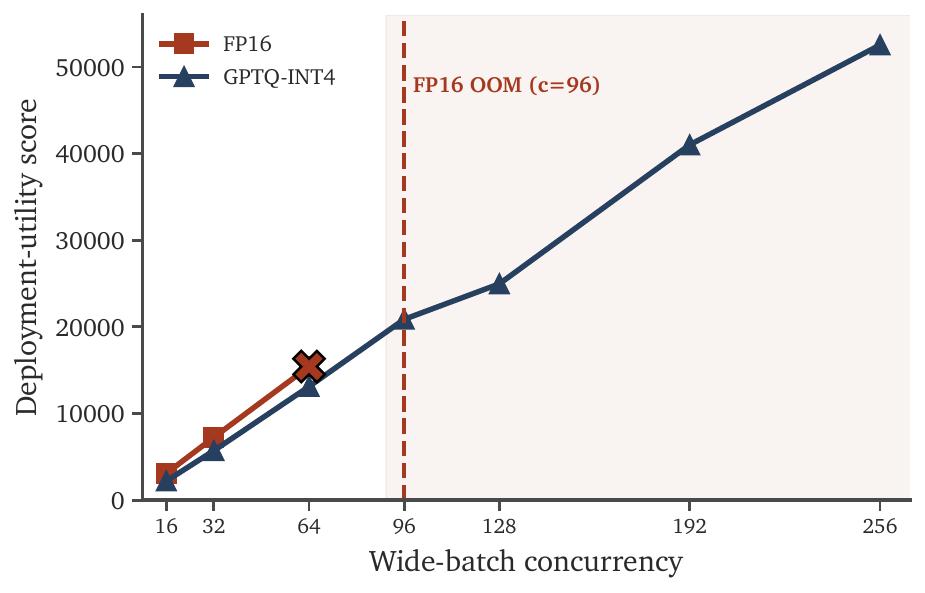}
\caption{Capacity cliff and INT4 capacity extension on the SwiftServe
stack. FP16 (terracotta) reaches a repeated memory cliff at the medium high
concurrency cell; INT4 (navy, GPTQ) continues through the rest of the
measured range. The vertical axis is the deployment utility score
defined and caveated in Section~\ref{subsec:capacity-protocol}. It is
not a standard throughput measurement. Detailed endpoints and
provenance appear in Tables~\ref{tab:phase-f}
and~\ref{tab:aggregate-throughput} and in
Appendix~\ref{appendix:data-sources}.}
\label{fig:capacity-cliff}
\end{figure}

Table~\ref{tab:phase-f} presents the full wide batch sweep for both the
reference and the quantized stack across the range we tested, and
Figure~\ref{fig:capacity-cliff} plots the resulting deployment utility
score.

\begin{table}[t]
\centering
\caption{Capacity sweep: HF-FP16 and GPTQ-INT4 at wide concurrency
($T=256$, SwiftServe HF stack with the upstream default quantized
kernel). FP16 fails at $c{=}96$ in all
three $n{=}3$ runs ($0/576$ requests succeed; consistent wall
$\approx 185$\,s as the SwiftServe HF server returns 500s for every
queued request after the OOM). GPTQ survives through $c{=}256$ with
$0/4{,}032$ failures across 4 cells and 3 runs.}
\label{tab:phase-f}
\small
\begin{tabular}{lrrrr}
\toprule
Concurrency & FP16 tok/s ($n{=}1$) & FP16 fail & GPTQ tok/s ($n{=}3$, $\pm$std) & GPTQ fail \\
\midrule
16  & 192.31  & 0/32   & 137.68 ($n{=}1$) & 0/32 \\
32  & 226.61  & 0/64   & 178.44 ($n{=}1$) & 0/64 \\
64  & 240.87  & 0/128  & 204.87 ($n{=}1$) & 0/128 \\
96  & \emph{0.00 (OOM, 3/3 runs)} & 576/576 & $217.37 \pm 3.18$ (CV $1.46\%$) & 0/576 \\
128 & ---     & ---    & $195.03 \pm 27.25$ (CV $13.97\%$) & 0/768 \\
192 & ---     & ---    & $213.60 \pm 2.05$ (CV $0.96\%$) & 0/1{,}152 \\
256 & ---     & ---    & $205.23 \pm 8.15$ (CV $3.97\%$) & 0/1{,}536 \\
\bottomrule
\end{tabular}
\smallskip

\noindent{\footnotesize \textit{Note:} GPTQ rows at $c\geq 96$ are
$n{=}3$ mean $\pm$ stddev (CV); GPTQ rows at $c\leq 64$ are $n{=}1$
single-run baselines from the earlier capacity measurements
(explicitly tagged $n{=}1$ in the table); FP16 rows at $c\leq 64$
are also $n{=}1$ single-run baselines from the earlier capacity
measurements, with the $c{=}96$ OOM cliff confirmed at $n{=}3$. The
historical $n{=}1$ rows are kept as a reference operating point but
should not be used for variance estimation.}
\end{table}

\begin{table}[t]
\centering
\caption{The vLLM-Marlin capacity sweep at $T{=}256$ (vLLM
runtime + Marlin INT4 kernel, $n{=}3$ each cell). All four cells
complete with $0/N$ failures; CV $\leq 4.55\%$ across runs. The
aggregate throughput is roughly flat across concurrency (the
single A5000 decode bandwidth ceiling), but the aggregate product
grows linearly with $c$.}
\label{tab:phase-g}
\small
\begin{tabular}{lrrl}
\toprule
Concurrency & Throughput (tok/s), $n{=}3$ mean $\pm$std (CV) & Failures & Status \\
\midrule
$c{=}96$  & $2{,}480.61 \pm 76.31$  ($3.08\%$) & $0/576$     & ok \\
$c{=}128$ & $2{,}636.47 \pm 119.99$ ($4.55\%$) & $0/768$     & ok \\
$c{=}192$ & $2{,}574.28 \pm 94.43$  ($3.67\%$) & $0/1{,}152$ & ok \\
\textbf{$c{=}256$} & $\mathbf{2{,}667.71 \pm 55.85}$ ($\mathbf{2.09\%}$) & $\mathbf{0/1{,}536}$ & ok \\
\bottomrule
\end{tabular}
\end{table}

\paragraph{Variance at one capacity workload.}
One of the quantized reference workloads varies more from run to run than
the others, because a single run was slower while the other two agreed.
All three completed every request, so this is scheduling jitter on the
host, and no failure or correctness problem. We report the mean
and spread and keep the slow run, and the per run details appear in
Appendix~\ref{appendix:70b-variance}.

\paragraph{Where the quantized stacks stand against the reference.}
On the reference serving stack, quantization alone is actually slower
than the unquantized version at every load where both run, trailing it by
a growing margin as the load rises. Its capacity advantage appears only
once the unquantized version can no longer run at all. The optimized
quantized stack reverses this. Where both run it serves about 10 times
as many tokens per second as the unquantized version at the same load,
and it also continues well past the point where the unquantized version
fails. It therefore both serves faster where the two overlap and extends
how many users can be served, unlike quantization on the reference stack.

\paragraph{The deployment utility score.}
We also report a deployment utility score for batch serving, defined and
caveated in Section~\ref{subsec:capacity-protocol}, which multiplies the
throughput by the number of concurrent users. Because it multiplies an
already aggregated throughput by the load, it is not itself a throughput
improvement; we use it only to combine service rate and reach into a
single number for the capacity comparison. The per workload values are in
Table~\ref{tab:aggregate-throughput}.

\begin{table}[t]
\centering
\caption{Bespoke deployment utility score
(\emph{aggregate token throughput product}: throughput $\times$
concurrency, defined in Section~\ref{subsec:capacity-protocol})
for FP16, HF-GPTQ ($n{=}3$ means at $c{\geq}96$;
$n{=}1$ single-run baseline at $c{=}64$), and vLLM-Marlin
($n{=}3$ means from the vLLM-Marlin capacity sweep). This score is not a standard throughput
measurement; it is reported only to compare configurations across
both throughput and feasible concurrency axes simultaneously.
HF-GPTQ at $c{=}256$ delivers $3.41\times$ FP16's best case score;
vLLM-Marlin at $c{=}256$ delivers $44.30\times$. The $13.00\times$
score ratio between HF-GPTQ and vLLM-Marlin combines both runtime
and kernel changes.}
\label{tab:aggregate-throughput}
\small
\begin{tabular}{llrl}
\toprule
Stack & Conc.\ & Aggregate product & Status \\
\midrule
FP16 (HF)            & $c{=}64$  & $240.87 \times 64 = 15{,}416$  & Peak (cliff at $c{=}96$, $n{=}3$) \\
\midrule
HF-GPTQ              & $c{=}96$  & $217.37 \times 96 = 20{,}867$  & --- \\
HF-GPTQ              & $c{=}128$ & $195.03 \times 128 = 24{,}964$ & --- \\
HF-GPTQ              & $c{=}192$ & $213.60 \times 192 = 41{,}011$ & --- \\
\textbf{HF-GPTQ}     & $\mathbf{c{=}256}$ & $\mathbf{205.23 \times 256 = 52{,}538}$ & $\mathbf{3.41\times}$ FP16 \\
\midrule
vLLM-Marlin          & $c{=}96$  & $2{,}480.61 \times 96 = 238{,}139$  & $15.45\times$ FP16 \\
vLLM-Marlin          & $c{=}128$ & $2{,}636.47 \times 128 = 337{,}468$ & $21.89\times$ FP16 \\
vLLM-Marlin          & $c{=}192$ & $2{,}574.28 \times 192 = 494{,}262$ & $32.06\times$ FP16 \\
\textbf{vLLM-Marlin} & $\mathbf{c{=}256}$ & $\mathbf{2{,}667.71 \times 256 = 682{,}934}$ & $\mathbf{44.30\times}$ FP16 \\
\bottomrule
\end{tabular}
\end{table}

On this score, quantization on the reference stack lands at about three
and a half times the best the unquantized version achieves, a gain that
comes entirely from reaching more users while running a little slower at
a fixed load. The optimized quantized stack lands far higher still, on
the order of 40 times, improving both the service rate and the reach
at once. The gap between the two quantized stacks reflects both a change
of runtime and a change of kernel, and we cannot assign it to the runtime
alone because both change together. The controlled runtime decomposition
in Section~\ref{subsec:controlled-rerun} applies at the highest load
where the intermediate baseline can still be measured; no such
intermediate is available at the higher loads, because it too runs out of
memory there. The two effects can be read separately. The first is reach,
where the unquantized version fails early while both quantized stacks
survive to the top of the range. The second is service rate at a fixed
load, where quantization on the reference stack is a little slower than
the unquantized version while the optimized stack is much faster and
continues past the point where the unquantized version fails. The score
combines the two into one number purely as a convenience.


\begin{figure}[t]
\centering
\includegraphics[width=0.85\linewidth]{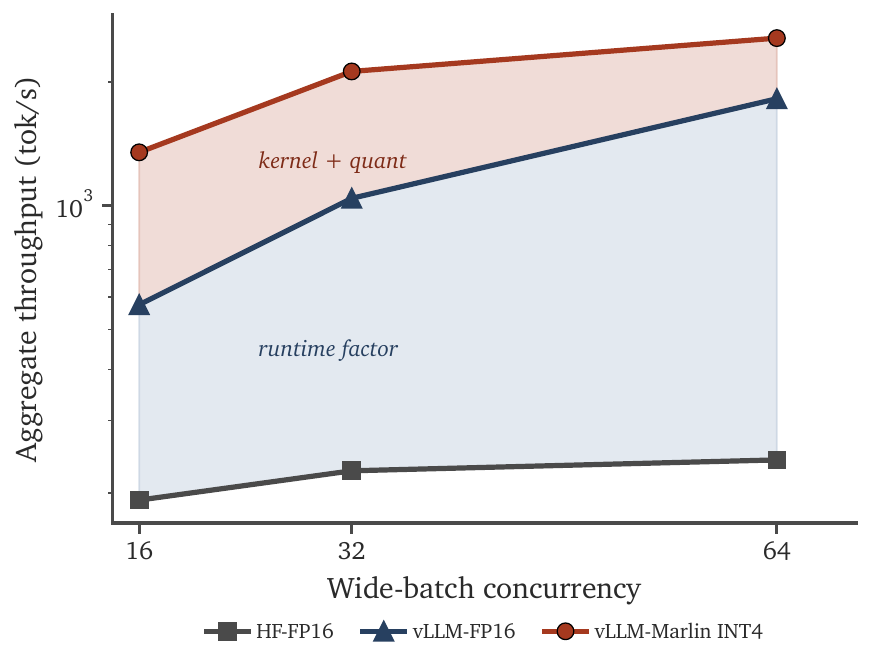}
\caption{Wide batch scaling on a single A5000 card. The matched FP16
intermediate (navy) separates the runtime change alone from the full
INT4 stack (terracotta). The gap between the baseline (gray) and the
intermediate widens as concurrency grows, so the runtime contribution
itself scales with concurrency (Section~\ref{subsec:attribution}).
Data provenance is in Appendix~\ref{appendix:data-sources}.}
\label{fig:throughput-concurrency}
\end{figure}


\begin{figure}[t]
\centering
\includegraphics[width=0.92\linewidth]{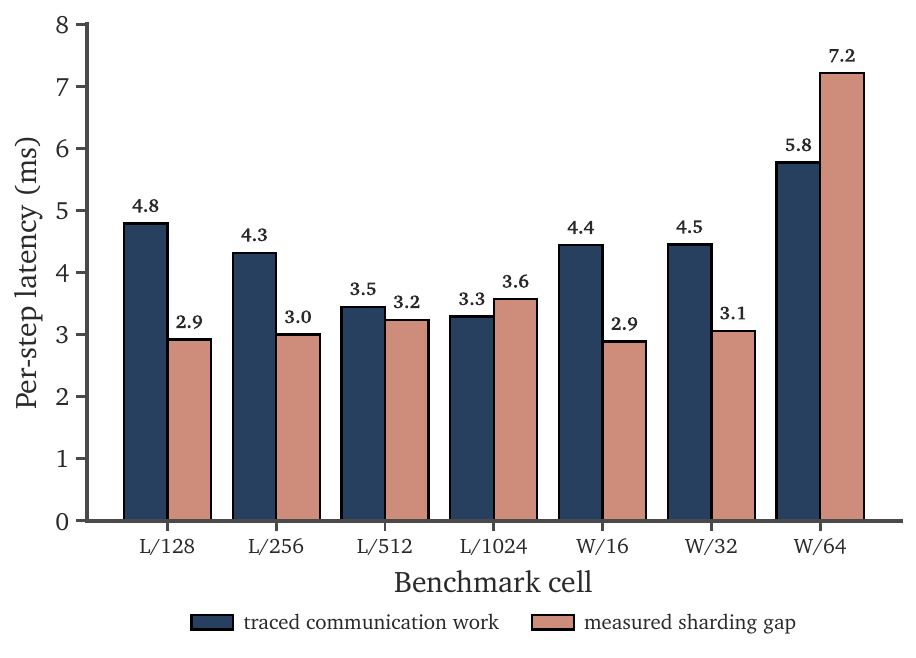}
\caption{Total traced communication work per output step (navy) versus
the measured sharding gap (light terracotta) across the concurrency and output
length sweep on the smaller model. At the wide batch operating point
(rightmost group) communication accounts for roughly 80\% of the
measured gap. In the other cells overlap with compute makes the
measured gap smaller than the total traced communication work. Full
numbers and the measurement protocol appear in
Appendix~\ref{appendix:tp-napkin} and Appendix
Table~\ref{tab:tp-napkin-cells}.}
\label{fig:tp-napkin-cells}
\end{figure}


\section{Discussion}
\label{sec:discussion}

\subsection{Why the Runtime Change Dominates at Wide Batches}

The advantage of the optimized runtime over the reference grows
several fold as the batch widens, from roughly double at the lightest
load to several times over at the widest.

\textbf{Hypothesis.}
At light load both runtimes keep the card reasonably busy, because the
reference packs its handful of requests into one pass and the matrix
multiply is large enough to run efficiently. As the batch widens, three
structural limits of the reference runtime compound:
\begin{enumerate}
  \item \emph{Wasted cache memory.}
        The reference reserves a contiguous padded block of cache for
        each batch, wasting memory on unused slots, which becomes
        substantial once the batch holds many requests of differing
        lengths.
  \item \emph{Rigid scheduling.}
        The reference groups requests into fixed batches and never
        refills the batch continuously the way vLLM
        does~\cite{yu2022orca, kwon2023vllm}, so when some requests
        finish early their slots sit empty until the whole batch
        completes.
  \item \emph{Per step overhead.}
        The reference generation loop carries per step overhead in the
        host language that the optimized runtime reduces.
\end{enumerate}
The optimized runtime manages its cache in small paged blocks and
refills the batch immediately, both of which keep the card busier at the
widest batch. Confirming how much each mechanism contributes would
require a profiler trace.

\subsection{Why the Kernel and Quantization Change Shrinks at Wide Batches}

The kernel and quantization factor falls from about two and a half times
on the decode heavy workloads to a little over one at the widest batch,
which runs against the naive expectation that a four bit kernel is always
about four times faster.

\textbf{Hypothesis.}
The quantized kernel delivers its full advantage when the work is
dominated by the matrix multiplies over the weights. On the decode heavy
workloads each step reads a large slice of the weights per token and the
matrix multiply is the dominant operation, so cutting the weight bytes to
a quarter and fusing the unpacking into the multiply produces a large per
step speedup, though less than fourfold because of the other per step
work.

At the widest batch the bottleneck shifts. The runtime now services many
requests at once, and attention, cache traffic, and scheduling take a
larger share of the time than the weight multiplies alone. The kernel
advantage is still real but is diluted into a smaller share of the total,
so the measured factor at the widest batch reflects that dilution.

One practical corollary follows. A team that cannot accept the small
quality cost of four bit weights should still adopt the faster runtime,
because in the controlled comparison the runtime change alone accounts
for about two thirds of the speedup at the widest batch. A team that can
accept the quality cost gains the kernel and quantization factor on top.

\paragraph{Controlled and operational attribution.}
The controlled rerun shows that the large runtime factor of the
production comparison is partly due to two effects outside the runtime
itself: a ceiling on how many requests the reference batches at once, and
the difference in sampling between the stacks. Removing both, the runtime
change accounts for about two thirds of the speedup at the widest batch,
and the kernel and quantization factor is essentially unchanged. We do
not treat either setting as the correct one. Production deployments
inherit the production setting, while controlled studies should use the
matched one.

\subsection{Why Sharding Does Not Stack}
\label{subsec:why-tp-fails}

Sharding a single instance across four cards raises throughput to only
about 1.5 times that of one card, far short of both the ideal of four
times and the practical hope of a doubling.

\paragraph{Two ways of measuring the per token gap.}
We estimate the per token latency at the widest batch, and the gap
between the sharded and single card versions, in two independent ways.
The first infers the latency from the measured throughput. The second
reads the serving library's own per token timing directly on a matched
rerun. The two agree closely, both placing the sharded version several
milliseconds faster per token than a single card, far short of the
fourfold gain four cards could give. The mechanism below is anchored on
the direct measurement.

\paragraph{Communication dominates the gap.}
A profiler trace on the same hardware measures how the sharded version
spends its time at the widest batch. Coordinating the shards accounts for
about four fifths of the per token gap there, which is the reverse of
what an earlier back of the envelope estimate had assumed. We extended
the trace to all seven workloads
(Appendix~\ref{appendix:tp-napkin}, Table~\ref{tab:tp-napkin-cells},
Figure~\ref{fig:tp-napkin-cells}). Across the sweep the coordination
time is roughly steady from workload to workload, and on most of them it
is actually larger than the measured gap, because it overlaps with
computation and only the part that does not overlap shows up in the gap.
The widest batch remains the cleanest single point to quote, because
there the computation is large enough that the gap exceeds the
coordination time.

Crucially, this cost is not explained by a slow link. A control on the
same hardware finds that the fast bridge and the slower link move the per
step data at essentially the same rate
(Appendix~\ref{appendix:nvlink-projection}), because at these small
message sizes the cost of each coordination step is dominated by a fixed
overhead to launch and synchronize it, and the speed of the link barely
enters. The full per call breakdown is in the appendix.

\paragraph{The resulting ordering.}
An earlier draft had assumed coordination was a small part of the gap and
that host contention and per shard inefficiency dominated. The
measurement reverses that ordering on this hardware: coordination is the
largest single part of the gap at the widest batch. A secondary but real
contributor is that each shard executes its matrix multiplications at
somewhat lower efficiency, since it works on a quarter of the problem. Host scheduling
and cache management make up the remainder. Because both the coordination
overhead and the non communication part of the gap are independent of the
link, replacing the slower link with the fast bridge would leave the
shortfall essentially where it is, far below the ideal.

\paragraph{Sharding gains during prompt processing do not carry through.}
An internal report finds a larger sharding advantage during prompt
processing alone (\cite{phase8_internal}, cited for context only). Our
full inference numbers are smaller because our workloads are decode heavy,
with a short prompt processed once and a long output generated one token
at a time. Prompt processing spreads a single coordination cost over a
great deal of computation, whereas each decode step pays that cost again
over much less computation, so the coordination weighs far more heavily
during decoding.

\paragraph{Sharding does not scale smoothly with load.}
The sharding gain is slightly smaller at one middle load than at the
widest. We suspect that at the middle loads the batch is not full enough
to hide the coordination behind useful computation, while at the widest
batch it is, which partly recovers the efficiency. This is a conjecture a
profiler study would settle.

\paragraph{The capacity outcome is not a runtime effect alone.}
The score gap between the two quantized stacks at the top of the range is
large, but it is not a runtime effect alone, because the two stacks
differ in both their runtime and their kernel. The decomposition that
isolates the runtime part elsewhere in the paper needs the intermediate
baseline, which cannot run at these high loads because it too runs out of
memory. We therefore report this gap as an operational difference between
two complete stacks, and note that isolating the runtime part at high
load would require a model small enough for the intermediate baseline to
fit, or a card with more memory.

\subsection{Why the Verdict Reverses on the Larger Model}
\label{subsec:70b-reversal}

The larger model comparison in Section~\ref{subsec:tp70b} reverses the
verdict the smaller model gave on the decode heavy workloads. We offer
the following mechanistic hypotheses, first for how the two instance
layout departs from simple scaling, and then for why it wins outright.

\paragraph{The decode heavy workloads fall well short of doubling.}
On these workloads the eight requests are split across two instances, so
each instance handles only four. With so few requests per instance the
cache is under used and the per token overhead on the host, such as
scheduling and data movement between the two halves of the machine, takes
a larger share of each token's time relative to the actual computation.
Simple scaling assumes each instance runs at the full eight request rate,
whereas each really runs at the slower four request rate.

\paragraph{The widest workload slightly exceeds doubling.}
At the widest batch a single instance is near the point where its cache
saturates and its throughput stops rising with load. Splitting into two
instances, each at half the load, places both in the more efficient part
of that curve, so the two together slightly exceed twice one instance. A
profiler trace would be needed to confirm exactly where that saturation
sets in.

\paragraph{Why the larger model reverses the verdict.}
The reversal is a firm empirical finding, and a profiler trace explains
it. On the larger model the weight matrices are big enough that the
quantized matrix multiplies dominate the time spent on each token, so
the fixed cost of keeping the shards in step matters less. Splitting the
four cards into two smaller instances also halves the number of cards
that must join each reduction, and with only two cards per group the
coordination can travel over a fast in memory path and bypass the
general purpose collective library. The single sharded instance cannot
use that path. Over the course of a run it issues on the order of two
hundred times as many explicit reduction calls as the two paired
instances combined, and it spends almost as much time coordinating as
computing, while the paired instances spend almost none. The penalty
comes from the sheer number of calls; their individual cost is low, since
each call is in fact cheaper on the single shard, where the payload is
divided among more cards; the paired instances simply avoid most of the
calls. Figure~\ref{fig:collective-collapse} and
Appendix~\ref{appendix:tp-napkin} give the counts. A control on the same
hardware confirms that the effect follows from how many cards take part;
the connection between them explains little, because moving each pair of instances
from the fast bridge to the slower link leaves the advantage largely in
place.
We profiled one workload for each configuration, so the trace supports
the mechanism at that point and is consistent with the throughput
reversal seen across all workloads, though we do not claim a separate
measurement at each one.

\paragraph{Comparison with the smaller model.}
On the smaller model the several instance layout at the widest batch
falls well short of simple scaling, whereas on the larger model the two
instance layout slightly exceeds it. The difference reflects the heavier
per token computation of the larger model, which reduces the relative
cost of contention on the host.

\subsection{Conditions Under Which Sharding Would Help}

The negative verdict on sharding the smaller model is conditioned on our
hardware and model, and it would reverse under the following conditions.

\begin{enumerate}
  \item \textbf{Lower coordination overhead.}
        At our per step message sizes the bottleneck is a fixed overhead
        to launch and synchronize each coordination step, which a control
        on the same hardware shows is about the same over the fast bridge
        and the slower link (Appendix~\ref{appendix:nvlink-projection}),
        so a faster link alone does not help. Sharding would benefit where
        that overhead is spread over larger transfers, at higher load or
        with larger models, or where all four cards share one low latency
        link through a switch fabric, which our pairwise bridges do not
        provide. The advantage of two instances on the larger model, by
        contrast, holds over both the fast bridge and the slower link, so
        it is not tied to any one link.

  \item \textbf{Models that do not fit on a single card.}
        The smaller quantized model fits comfortably on one card, so
        sharding is optional. The larger model does not fit on one card,
        so some form of splitting is required. In that case the result in
        Section~\ref{subsec:tp70b} applies directly, because both layouts
        use all four cards and the choice is between more instances and a
        wider shard, since splitting is already required.

  \item \textbf{Latency sensitive single request work.}
        Our workloads measure aggregate throughput at eight or more
        concurrent requests. For a single request needing the lowest
        possible per token latency, sharding can lower that latency by
        splitting the computation across cards. We do not measure this
        case.
\end{enumerate}

\subsection{Threats to Validity and Limitations}

We recap and expand the limitations introduced earlier:

\begin{enumerate}
  \item \textbf{Single hardware configuration.}
        All experiments use four A5000 cards. A control on the same
        hardware (Appendix~\ref{appendix:nvlink-projection}) bounds how
        much the interconnect matters within this class of card, but the
        results may not carry over to newer card generations, to a machine
        where all four cards share one link through a switch fabric, or to
        setups that offload work to the host.

  \item \textbf{Model coverage.}
        The stacking, capacity, and smaller model routing results cover
        the smaller model, and the larger model comparison covers the
        larger one. The reversal on the larger model depends on the
        hardware and holds across the interconnect within this class of card,
        and it may not carry over to other models or card generations. The
        attribution decomposition, once measured on a single model, is now
        checked at the headline workload on three architecturally distinct
        models of roughly seven to eight billion parameters, where the
        kernel and quantization factor agrees to within about one and a
        half percent (Section~\ref{subsec:cross-model}). We therefore
        treat the stability of that factor across models as a measured
        finding on this hardware, and no longer an open question. We do not
        measure the runtime part again on the other two models, because the
        reference stack would need porting to their architectures, so the
        cross model evidence bounds only the kernel and quantization part.
        Whether the decomposition holds for very different or much larger
        models is outside our scope.

  \item \textbf{Single quantization method.}
        All quantization experiments use one four bit method. Other
        methods (\cite{lin2024awq}, eight bit floating point, and others)
        would produce different kernel factors.

  \item \textbf{No four card shared link; only pairwise links measured.}
        Our control compares two cards on the fast bridge against two
        cards on the slower link
        (Appendix~\ref{appendix:nvlink-projection}). At our per step
        message sizes the two move data at essentially the same rate, and
        the two instance throughput on the larger model matches closely
        across the two. Because a fixed overhead dominates the coordination,
        and link speed has little influence, a faster link leaves both the
        coordination cost and the large non communication part of the gap
        unchanged, holding the stacking factor where we measured it. What
        we cannot measure is all four cards sharing one link, since the
        bridges on these cards are pairwise and a four card shared link
        would require a switch fabric found only on data center cards.
        Whether sharding stacks under such a fabric, and how it behaves on
        other card generations, are the most valuable follow ups.

  \item \textbf{Different sampling between the stacks.}
        The reference cells use deterministic greedy decoding while the
        optimized cells use random sampling (Section~\ref{subsec:sampling}).
        The argument that the difference in cost between the two is small
        relative to the matrix multiplies is plausible and awaits direct
        validation, and a future revision should match the two.

  \item \textbf{Small synthetic prompt pool.}
        We use a small pool of synthetic prompts, each about a dozen
        tokens long. Generalizing to realistic prompt distributions with
        longer and more variable prompts is left for future work, since
        the throughput trends may differ at the extremes of prompt length.

  \item \textbf{Sample sizes vary by experiment.}
        The original attribution decomposition rests on a single run per
        workload, so we report no variation for those. The stacking sweep,
        the smaller model several instance comparison, and both larger model
        comparisons rest on five runs each, with low run to run variation
        everywhere except the single sharded baseline, which is the
        noisiest. The capacity cliff rests on three runs. One capacity
        workload varied more from run to run because one run was slow, and
        we report its mean and spread and keep the slow run in.
        The decomposition itself is an exact identity
        (Eq.~\ref{eq:product-law}), so any tiny discrepancy from
        multiplying the rounded table values is rounding, and never
        evidence that the factors are independent.

  \item \textbf{Uncontrolled clock and thermal state.}
        We did not pin the card clocks, did not enforce idle time between
        workloads, and did not control for heat building up over sustained
        batches.

  \item \textbf{Several instance scaling is not linear and depends on the
        workload.}
        Measuring the several instance layout directly
        (Table~\ref{tab:multiinstance}) shows that it recovers only a
        fraction of what simply adding up four separate cards would
        suggest. We attribute this to contention for the shared host,
        since the instances compete for one processor, one memory bus, and
        one storage channel, and we expect the efficiency to rise with
        load because at high load each instance is busy enough to hide the
        contention behind useful computation. The way the outcome depends
        on the workload is the load bearing routing
        implication for practitioners. On the smaller model a single
        sharded instance wins the decode heavy workloads while several
        independent instances win the wide ones, and at the largest wide
        workload the single shard runs out of memory, which leaves the
        independent instances the only option. On the larger model the
        independent instances win throughout, on a baseline that itself
        runs without trouble. This may not generalize to settings where
        the instances do not share a host.

  \item \textbf{Reversal mechanism on the larger model.}
        Several independent instances beat a single sharded instance on
        every workload of the larger model, and the reason is that fewer
        cards take part in each reduction; the link between the cards
        plays a minor role. The heavier per token compute of the larger
        model leaves less room for coordination cost, and with only two
        cards per group most of the coordination moves onto a fast in
        memory path. A control on the same hardware, in which the
        instances are moved from the fast bridge to the slower link,
        leaves the advantage in place
        (Appendix~\ref{appendix:nvlink-projection}). The single shard is
        also the noisier configuration, which is consistent with its
        larger set of communicating cards amplifying the host side timing
        jitter that smaller groups absorb.

  \item \textbf{Only two runtimes compared.}
        We tried to add two further serving runtimes as third and fourth
        points of comparison, but both were blocked by infrastructure
        incompatibilities on our cluster, which we document in
        Appendix~\ref{appendix:cross-runtime-attempts}. Adding them would
        not change the algebraic split for the three stacks we measured,
        though it would test whether the runtime conclusion generalizes to
        other runtimes.

  \item \textbf{Controlled and operational runtime factors are not reconciled.}
        The controlled rerun is complete across the full sweep of seven
        cells (Section~\ref{subsec:controlled-rerun}). Its runtime figure
        and share differ materially from the production numbers because
        the two settings differ in batching regime and sampling mode, and
        we report both as two distinct deployment questions. We do not run
        a factorial ablation that isolates the batching change from the
        sampling change, so the exact split between those two confounds
        remains open.
\end{enumerate}


\section{Conclusion}
\label{sec:conclusion}

We have presented a decomposition study of language model inference
throughput on four A5000 cards, drawn from a single host and serving an
eight billion parameter model. The cards are bridged in pairs, and a
control on the same hardware indicates that the connection between them
is unlikely to be the operative bottleneck at our message sizes
(Appendix~\ref{appendix:nvlink-projection}). By adding an intermediate
baseline that keeps the faster runtime and drops the quantized kernel, we
split the large speedup of the fully optimized stack over the reference
into a runtime part and a kernel and quantization part, whose product is
the whole by construction. This split is a bookkeeping device, and it
makes no claim that the two parts act independently.

Our headline attribution comes from a matched controlled measurement: at
the wide batch operating point the runtime change accounts for about two
thirds of the overall speedup measured on a logarithmic scale, and the
kernel and quantization change accounts for the rest. As a deployment
case study we also report what a direct production swap would show,
where the apparent runtime share is much larger. That gap reflects
differences in batching and in sampling between the two stacks, well
beyond the runtime alone, and we do not turn either figure into a causal statement,
because the intermediate cancels by construction and the two stacks
differ in more than their runtime. Under the production setting the
runtime share grows as the batch widens.

Sharding a single instance across all four cards raises throughput to
only about 1.5 times that of one card, far below a practical doubling
threshold and further still from perfect scaling. A direct latency
measurement and a profiler trace attribute most of the shortfall to the
time spent coordinating the shards, with a smaller but still meaningful
part coming from each shard executing its matrix multiplications at
somewhat lower efficiency and from host side scheduling. A control on the same
hardware shows that the faster of the two available links does not close
the gap, so the shortfall is not an artifact of the slower link. A
four way arrangement that keeps all cards on the fast link would need a
switch fabric we do not have, so it remains untested and could still
help.

Whether to run one sharded instance or several independent ones depends
on the workload. Assuming that several instances simply scale linearly
overstates their throughput, most of all on the decode heavy workloads,
where on the smaller model a single sharded instance actually wins. On
the wide workloads several instances win, and at the largest one they
are the only option because the single shard runs out of memory. On the
larger model the several instance layout wins throughout.

Finally, the capacity outcome depends on both the runtime and the
kernel. Quantization roughly quadruples the number of concurrent users a
card can sustain before it runs out of memory, a limit the reference
stack reaches well before either quantized stack. On a deployment utility score
that rewards serving more users at once, both quantized stacks far exceed
the reference. For one of them the entire gain comes from serving more
users, since at a fixed load it delivers no faster service; the other is
faster at a fixed load as well. The exact scores mix the runtime and the
kernel changes together and are reported in the results as an operational
gap across the whole stack, and never as a contribution of the runtime
alone.

\paragraph{Observed best configuration on the tested hardware.}
Table~\ref{tab:routing} summarizes the configuration that performed best
at each level of load on the four card machine tested here, under our
small synthetic prompt pool. Every row is conditioned on this hardware
and workload and on the documented difference in sampling between the
stacks. The attribution rows rest on three runs each, apart from the
production case study, which is a single run; the smaller model stacking
and routing rest on five runs each, the larger model comparison on the
matched three run control, and the capacity cliff on three runs.

\begin{table}[t]
\centering
\caption{Observed best configuration at each level of load on the
four card machine tested here (single node $4{\times}$ A5000
24\,GiB PCIe gen4, NVLink-bridged pairs, ten prompt synthetic
workload). Sample sizes and the sampling mode asymmetry between the
stacks are documented in the paragraph above.}
\label{tab:routing}
\small
\begin{tabular}{p{2.7cm}p{3.0cm}p{5.4cm}}
\toprule
Scenario & Observed best config. (this hardware) & Rationale (with caveats) \\
\midrule
Single-GPU 8B serving
  & vLLM-Marlin INT4 (or vLLM-FP16 if no INT4 quality budget)
  & Headline (controlled, matched-greedy $n{=}3$): $2.58\times$ end-to-end at $c{=}64$, runtime-arm log-share $67.9\%$. Operational case study ($n{=}1$, batch ceiling 8, ancestral vLLM): $10.62\times$ end-to-end, log-share $86\%$ (Table~\ref{tab:c1-sensitivity}). \\
Multi-GPU 8B routing
  & TP=4 for $c{=}8$ long-output; multi-instance for $c{\geq}32$ wide-batch (and the only viable backend at $c{=}128$, $7{,}244$\,tok/s aggregate)
  & TP=4 wins long\_c8 by $1.28$--$1.30\times$ ($n{=}5$); multi-instance wins wide-batch up to $1.46\times$ at $c{=}64$ ($n{=}5$); crossover between $c{=}16$ (TP=4 by $1.16\times$) and $c{=}32$ (multi-instance by $1.11\times$). \\
70B (model does not fit on a single GPU)
  & $2{\times}$TP=2 multi-instance
  & On the larger model two smaller instances beat a single sharded instance on every workload, by about seven percent, and the advantage largely holds even when the two instances use only the slower link, so it does not come from the interconnect (Appendix~\ref{appendix:nvlink-projection}). On the smaller model, sharding one instance across four cards adds well under a doubling and stands only as a lower bound. \\
High concurrency on one card ($c{\geq}96$)
  & vLLM-Marlin INT4
  & FP16 runs out of memory at $c{=}96$ ($3/3$ runs). Both INT4 stacks extend max sustainable concurrency to $c{=}256$ ($n{=}3$); on the bespoke utility score (\S\ref{subsec:capacity-protocol}) HF-GPTQ scores $3.41\times$ and vLLM-Marlin $44.30\times$. A control on the same hardware (Appendix~\ref{appendix:nvlink-projection}) measures similar realized reduction bandwidth at the per step payload ($0.95\times$, single run), so a bandwidth explanation is unlikely (an arrangement placing all four cards in a single fast switch fabric is untested). \\
\bottomrule
\end{tabular}
\end{table}

These observations apply only to our specific hardware and workload.
Systems that place all four cards in a single fast switch fabric, longer
prompt distributions, or strict latency service targets may invert any of
these orderings.

\paragraph{Methodological takeaway.}
The broader lesson is methodological. Reporting a whole stack speedup
without an intermediate baseline conflates the runtime, kernel, and
quantization contributions in a way that cannot be separated after the
fact. Future serving benchmarks should include at least one intermediate
baseline that holds the runtime fixed while changing the kernel, or the
reverse, so that the contribution of each part is defensible.

\bibliographystyle{plainnat}
\bibliography{refs}

\appendix

\section{Prompt Pool}
\label{appendix:prompts}

The complete pool of 10 prompts used by all benchmark scripts is
reproduced verbatim below. The same list is shared between the reference
harness and the benchmark scripts for the optimized stack, and the
prompts are cycled deterministically in a fixed order. The exact script
locations are catalogued in Appendix~\ref{appendix:data-sources}.

\begin{enumerate}
  \item \emph{Write a short story about a robot learning to paint.}
  \item \emph{Explain the difference between TCP and UDP in detail with examples.}
  \item \emph{Describe how a large language model is trained, from data to deployment.}
  \item \emph{What are the key trade-offs between SQL and NoSQL databases?}
  \item \emph{Outline the steps to design a RESTful API for a todo-list application.}
  \item \emph{Compare and contrast bubble sort, merge sort, and quicksort.}
  \item \emph{Explain how garbage collection works in modern programming languages.}
  \item \emph{Write a tutorial on how to use Git rebase versus merge.}
  \item \emph{Describe the architecture of a typical microservices deployment.}
  \item \emph{What is dynamic programming and how does it differ from greedy algorithms?}
\end{enumerate}

The prompts average approximately $12$ tokens under the Llama 3.1
tokenizer. They are intentionally short so that each workload is dominated
by generating the output rather than by processing the prompt, since the
output lengths run from $128$ to $1{,}024$ tokens. We do not separately
measure how the time splits between processing the prompt and generating
the output.

\section{Variance Details for the Larger Model and a Capacity Workload}
\label{appendix:70b-variance}

Across the five runs of the single sharded larger model, the run to run
variation on the four decode heavy workloads is $7.25$, $10.65$, $17.86$,
and $12.65$ percent, and on the three wide workloads it is $13.41$,
$10.82$, and $1.04$ percent. The widest batch, at about one percent, is
the one quiet workload, because the computation there is large enough to
saturate the cards. The noise comes mostly from a split in the runs: on
several decode heavy workloads runs one through three cluster high while
runs four and five drop by about a quarter, whereas the widest batch does
not show this split. We read the asymmetry as support for the idea that
coordinating more cards amplifies timing jitter on the host, though a
profiler trace would be needed to confirm the mechanism.

\paragraph{The one noisy capacity workload.}
On the quantized reference stack at a load of 128
concurrent users, the three runs measured $208.92$, $163.64$, and
$212.53$ tokens per second; only the middle run fell below 200.
All 256 requests succeeded in every run. We attribute
the slow run to scheduling jitter on the shared machine rather than to a
memory failure or a correctness problem.

\section{Communication in the Sharded Configuration: Estimate Versus Measurement}
\label{appendix:tp-napkin}

This appendix works through two accounts of the per token latency gap
between the sharded configuration and a single card: a closed form back
of the envelope estimate, kept for intuition, and a direct profiler
measurement on the same hardware, which is the primary evidence. The two
disagree by roughly a factor of four on the share attributed to
communication, and we rely on the measurement.

\paragraph{The back of the envelope estimate.}
For the smaller model, with a hidden width of $4{,}096$ in half
precision, one common simplification treats the data each card sends at
every reduction as $4{,}096 \times 2\,\text{bytes} = 8\,\text{KiB}$ per
token, assuming a single request in flight. Under that simplification and
a link running at about $32$\,GB/s, the transfer time would be
$8{,}192 / 32 \times 10^9 \approx 0.26\,\mu$s, plus a fixed overhead of
about $5\,\mu$s to launch each collective, for a total of about
$5.3\,\mu$s per call. A transformer block has two reductions per layer,
one after the attention output and one after the feed forward output, so
across the model's layers there are about $2 \times 32 = 64$ reductions
per token. The single request estimate would then predict a
communication contribution of about $5.3 \times 64 \approx 0.34$\,ms per
token.

This single request simplification is wrong for our setting, where many
requests are served at once. The tensor reduced at each step has one row
per concurrent request, so at the widest batch each card sends
$64 \times 4{,}096 \times 2 = 512$\,KiB per reduction rather than
$8$\,KiB. The transfer time grows accordingly to about $16\,\mu$s per
reduction, the total per call rises to about $21\,\mu$s, and the
corrected communication contribution is
$\approx 21\,\mu\text{s} \times 64 = 1.34\,\text{ms}$ per decode step.

\paragraph{Direct profiler measurement.}
A profiler trace at the widest batch on the same four card hardware,
over a roughly 60 second steady state window, directly measures the
wall time of each communication call and the share of each decode step
spent communicating. The communication kernels logged in that window
break down by class as follows, with averages taken from the trace
export:
\begin{center}
\small
\begin{tabular}{lrr}
\toprule
Communication kernel & avg wall time / call & total ms / share \\
\midrule
Sum reduction across cards   & $741\,\mu$s & $2{,}120$\,ms ($22.1\%$) \\
Point to point send/receive  & $523\,\mu$s & $5{,}912$\,ms ($61.6\%$) \\
Broadcast                    & $15\,\mu$s  & $1{,}561$\,ms ($16.3\%$) \\
\bottomrule
\end{tabular}
\end{center}
The measured cost of a sum reduction, about $741\,\mu$s per call, is some
35 times the $21\,\mu$s estimate, and the point to point
transfers, which the estimate did not model at all, contribute more total
time than the sum reductions ($61.6$ versus $22.1$ percent of the total).
Aggregating all three classes over the window gives $9{,}594$\,ms of
communication time. Dividing by the number of decode steps gives a
communication cost per step of $3.5$ to $5.3$\,ms, where the lower end
uses a throughput based step count of about $2{,}718$ steps and the upper
end uses a communication kernel count based step count of about $1{,}811$
steps.

\paragraph{Reconciling the estimate with the measurement.}
At the widest batch the communication time per step, which varies from
$5.30$ to $5.77$\,ms across two captures, accounts for $74$ to $80$
percent of the directly measured per token gap of $7.17$ to $7.21$\,ms,
taken from the serving library per token timing on a matched three run
rerun of the same workload with graph capture on
(Section~\ref{subsec:why-tp-fails}). To test whether this share is
particular to that workload or general, we extended the trace to the
other six workloads (results in Table~\ref{tab:tp-napkin-cells} below).
The communication time per step stays in a narrow band of $3.3$ to
$5.8$\,ms across all seven, while the measured gap varies more widely,
from $2.9$ to $7.2$\,ms. On six of the seven the communication time
exceeds the gap, which means the communication overlaps heavily with
computation and the gap captures only the part that does not overlap. The
widest batch remains the cleanest single point to quote because there the
computation is large enough that the gap exceeds the communication time.
The part of the gap that is not communication, about $1.9$ to $3.7$\,ms
per step, is consistent with each shard running its matrix multiplies a
little less efficiently: the same trace records $1.23$ times more total
multiply time under sharding despite a $4.4$ times higher kernel count,
so each shard works on a smaller and less efficient problem. The rest is
scheduling and cache management on the host.

\paragraph{Per workload communication breakdown.}
Table~\ref{tab:tp-napkin-cells} reports the communication breakdown for
each of the seven workloads.

\begin{table}[h]
\caption{Per cell communication breakdown on the smaller model under
four way tensor parallelism. Detailed measurement protocol and exact
metric labels appear in the table and surrounding appendix text.}
\label{tab:tp-napkin-cells}
\begin{center}
\small
\begin{tabular}{lrrrrrrr}
\toprule
Cell ($c$, $T$) & ITL TP=4 & ITL single & gap & NCCL/step & NCCL/gap & AR avg & SR avg \\
                & (ms)     & (ms)        & (ms)& (ms)      & (\%)     & ($\mu$s)& ($\mu$s)\\
\midrule
long  $c{=}8,\,T{=}128$  & 7.04  &  9.96 & 2.92 & 4.79 & 164 & 741 &  55 \\
long  $c{=}8,\,T{=}256$  & 7.07  & 10.06 & 3.00 & 4.32 & 144 & 867 &  61 \\
long  $c{=}8,\,T{=}512$  & 7.04  & 10.28 & 3.24 & 3.45 & 107 & 939 &  59 \\
long  $c{=}8,\,T{=}1024$ & 7.17  & 10.74 & 3.57 & 3.29 &  92 & 735 &  59 \\
wide  $c{=}16,\,T{=}256$ & 7.80  & 10.69 & 2.89 & 4.44 & 154 & 758 &  90 \\
wide  $c{=}32,\,T{=}256$ &10.27  & 13.33 & 3.06 & 4.45 & 145 & 627 & 169 \\
wide  $c{=}64,\,T{=}256$ &14.61  & 21.82 & 7.21 & 5.77 &  80 & 522 & 321 \\
\bottomrule
\end{tabular}
\end{center}
\end{table}

\paragraph{Per arm 70B communication breakdown.}
Table~\ref{tab:tp-napkin-70b} reports the profiler comparison on the
larger model at the wide batch operating point, and
Figure~\ref{fig:collective-collapse} visualizes the two headline
contrasts: the roughly $192\times$ gap in explicit collective count and
the $82.0\%$ versus $1.7\%$ communication-to-compute ratio.

\begin{table}[h]
\caption{Communication comparison on the larger model at the wide batch
operating point. The four way arm spends much more time in
communication and issues many more explicit collective calls than two
smaller sharded instances. Exact counts and ratios appear in the table.}
\label{tab:tp-napkin-70b}
\begin{center}
\small
\begin{tabular}{lrrrrr}
\toprule
Arm & AR calls & AR avg ($\mu$s) & NCCL total (ms) & Marlin total (ms) & NCCL/Marlin (\%) \\
\midrule
TP=4 (4 GPUs)            & $494{,}592$ & $181.8$    & $91{,}421$ & $111{,}470$ & $82.0$ \\
$2{\times}$TP=2 inst$_0$ &   $1{,}288$ & $947.2$    &  $1{,}518$ &  $91{,}745$ &  $1.7$ \\
$2{\times}$TP=2 inst$_1$ &   $1{,}288$ & $1{,}015.8$&  $1{,}615$ &  $92{,}124$ &  $1.8$ \\
\midrule
$2{\times}$TP=2 sum      &   $2{,}576$ & n/a        &  $3{,}133$ & $183{,}869$ &  $1.7$ \\
\bottomrule
\end{tabular}
\end{center}
\end{table}

\paragraph{Reproducibility.}
Trace artifacts and extraction scripts are shipped with the code
release; the table reports the resulting per class summaries. Detailed
artifact sizes, file paths, and sampling protocols appear in
Appendix~\ref{appendix:data-sources}.

\paragraph{Why we retain the estimate.}
The estimate still serves a pedagogical function: it makes the
algebraic structure of tensor parallel collective cost transparent
(per rank payload times wire bandwidth, plus launch overhead, times
collectives per step). Its quantitative prediction was wrong on this
hardware essentially because both inputs were optimistic. The
realized algorithmic bandwidth at the relevant payload size is about
five times below the nominal value used by the estimate (measured by an
independent four rank microbenchmark on the same hardware), and the
estimate omits one of the two collective categories entirely. The
measurement is therefore not surprising; it is the price of replacing
nominal numbers with realized ones.

\section{Data Sources and Artifact Paths}
\label{appendix:data-sources}

All numerical claims in the main text are reproducible from JSON
artifacts shipped with the code release. Table~\ref{tab:data-sources}
catalogs them; paths are relative to the repository root. The controlled
C1 rerun was produced by the wrapper script
\path{scripts/decomposition_greedy_n3.sh} at commit \texttt{a07164d};
the operational reference arm ran through the SwiftServe wrapper with
batched decoding enabled at an eight request ceiling.

\begin{table}[h]
\centering
\caption{Artifact catalog. One row per measurement group; asterisks are
shell wildcards over backend and cell names.}
\label{tab:data-sources}
\small
\begin{tabular}{@{}p{5.1cm}p{9.6cm}@{}}
\toprule
Measurement group & Artifact path(s) \\
\midrule
Operational C1, reference arm ($n{=}1$)
  & \path{benchmarks/results/p11b_fp16_*.json} \\
Operational C1, intermediate arm
  & \path{benchmarks/results/phase12_5_vllm_fp16_*.json} \\
Operational C1, full stack arm
  & \path{benchmarks/results/phase12_marlin_*.json} \\
\addlinespace
Controlled C1, headline cell ($n{=}3$; 9 per-run files, 3 backends $\times$ 3 runs)
  & \path{benchmarks/results/decomposition_greedy_*_run{0,1,2}_c64_mt256.json} \\
Controlled C1, headline summary
  & \path{benchmarks/results/decomposition_greedy_n3_summary.json} \\
Controlled C1, envelope cells (54 per-run files, 3 backends $\times$ 3 runs $\times$ 6 cells; Table~\ref{tab:c1-envelope})
  & \path{benchmarks/results/decomposition_greedy_envelope_*_run{0,1,2}_*.json} \\
Controlled C1, envelope summary
  & \path{benchmarks/results/decomposition_greedy_envelope_summary.json} \\
\addlinespace
8B stacking and single GPU baselines ($n{=}5$)
  & \path{benchmarks/results/phase13_multirun_*.json} \\
8B multi-instance ($n{=}5$; wrapper reproduces Table~\ref{tab:multiinstance})
  & \path{benchmarks/results/phase12_7b_multi_instance_nodocker_*.json},
    \path{scripts/multi_instance_nodocker_bench.sh} \\
70B sharding and multi-instance ($n{=}5$)
  & \path{benchmarks/results/phase13_5_*.json} \\
\addlinespace
FP16 and HF-GPTQ capacity sweep ($n{=}3$ at $c{\geq}96$)
  & \path{benchmarks/results/phaseF_*.json} \\
vLLM-Marlin capacity sweep ($n{=}3$)
  & \path{benchmarks/results/phaseG_vllm_marlin_capacity_*.json} \\
\addlinespace
Verdict documents
  & \path{docs/phase12_5_attribution_verdict.md},
    \path{docs/phase12_6_tp_verdict.md},
    \path{docs/phaseF_cliff_verdict.md},
    \path{docs/phase13_5_70b_multi_instance_verdict.md} \\
\bottomrule
\end{tabular}
\end{table}

\section{A Matched NVLink Versus PCIe Control on the Tested Node}
\label{appendix:nvlink-projection}

An earlier draft projected the effect of a faster interconnect by
algebra alone, under the assumption that NVLink hardware was unavailable
on this host. Direct inspection showed the assumption to be incorrect:
the host bridges its cards into NVLink pairs, so the two cards within a
pair reach each other over NVLink while separate pairs reach each other
over PCIe. This lets us compare the two directly by pinning a pair of
cards to NVLink or to PCIe. We report that direct measurement here, and
it replaces the earlier projection, which assumed the collective was
limited by bandwidth and so overstated the gain from sharding.

\paragraph{The two links are close at decode message sizes.}
A synthetic reduction sweep between two cards on an idle node, with the
pair pinned to the fast bridge and then to the slower link so that the
cards are identical and only the connection changes, reports the realized
bandwidths in Table~\ref{tab:nvlink-measured} (single run per message
size). At the $512$\,KiB message size that matches the widest batch, the
ratio of the two links is $0.95$, within about five percent, and across
the full $8$\,KiB to $1$\,MiB sweep it stays between $0.81$ and $1.01$,
straddling one. In this range both links sit on a common floor of roughly
100 microseconds per call, so the fast bridge's nominal $112$\,GB/s
advantage does not show up: at these small message sizes the collective
is dominated by the fixed cost of launching and synchronizing each call
rather than by transfer time. Turning off the direct card to card path on
the fast bridge slows the collective by a factor of $1.73$ at
$512$\,KiB, and the slower link with its direct path on reaches a
bandwidth close to the fast bridge, $5.52$ against $5.27$\,GB/s. We did
not run the slower link with its direct path turned off, so we report
these as measured points rather than a full separation of transfer time
from copy cost.

\begin{table}[h]
\centering
\small
\begin{tabular}{lr}
\toprule
Configuration ($2$-rank allreduce, $512$\,KiB, single run) & Realized algbw \\
\midrule
NVLink pair, peer transport on            & $5.27$\,GB/s \\
PCIe (NODE) pair, peer transport on       & $5.52$\,GB/s \\
\textbf{NVLink / PCIe ratio}              & $\mathbf{0.95\times}$ \\
NVLink pair, peer transport disabled      & $3.05$\,GB/s \\
Four-way, two NVLink islands (TP=4 shape) & $8.06$\,GB/s \\
\bottomrule
\end{tabular}
\caption{Control comparing the fast bridge against the slower link on
idle cards, one run per configuration. At the message size matching the
widest batch the two are within about five percent; the small collective
sits on a fixed floor per call, so the fast bridge's nominal bandwidth
advantage does not show up.}
\label{tab:nvlink-measured}
\end{table}

\paragraph{The advantage does not depend on the interconnect.}
We ran all three configurations of the larger model comparison in one
session on one idle group of four GPUs, using the same server and
sampling and repeating each workload three times: a single sharded
instance, two instances with each on the fast bridge, and two instances
with each on the slower link (Section~\ref{subsec:tp70b},
Table~\ref{tab:tp70b}). Moving the two instances from the fast bridge to
the slower link changes their throughput by about two percent on
average, so throughput is essentially unchanged across the two
connections. Because
the same session also measured the single sharded baseline, this
comparison gives the advantage directly rather than through a baseline
borrowed from another run: about seven percent on the fast bridge and
about five percent on the slower link. It replaces an earlier estimate
of roughly 15 percent, which had used a single instance baseline
from a separate session that happened to run several percent slower on
the wide workloads. The advantage comes from having fewer cards join
each reduction, not from the link between them.

\paragraph{Bearing on the sharding result.}
Together these controls constrain the finding that sharding does not
stack, without directly running the smaller model sharded path again
under a matched interconnect, which we did not do. The microbenchmark
shows the fast bridge and the slower link move data at the same realized
rate at the relevant message size, and the larger model comparison shows
that swapping one for the other does not change the several instance
throughput. The profiler account further attributes a large share of the
sharding gap to work other than communication, namely each shard running
its matrix multiplies a little less efficiently and the extra scheduling
on the host, none of which any change of link affects, and it shows the
communication spread over many small launches rather than concentrated in
transfer time. Together these make a bandwidth explanation of the
shortfall unlikely and suggest a faster link alone would not lift the
stacking factor materially above the measured $1.49$ times, in contrast
to the earlier projection of $2.33$ times, which assumed the collective
was limited by bandwidth.

\paragraph{What remains unmeasured.}
Putting all four cards on one shared fast link would require a switch
fabric; the bridges on these cards are pairwise only, so all four cannot
share a single fast domain, and we cannot rule out that such a domain
would let sharding stack better. The pairwise controls above reduce the
chance that the bandwidth of the pair link explains the result, but they
do not substitute for a measurement with all four cards on one shared
fast link, which remains the most valuable follow up.

\paragraph{Reproducibility and sample size.}
The driver and the result files for these controls are catalogued with
the code release. The microbenchmark is a single run per message size and
its ratios are sensitivity checks; the larger model control is the
three run matched comparison of all three configurations in one session
on one island, with the median reported for each workload. The direct
latency measurement and the profiler inputs are as in
Appendix~\ref{appendix:tp-napkin}.

\section{Setup Attempts for Two Further Runtimes}
\label{appendix:cross-runtime-attempts}

This appendix records why the two further runtimes we tried could not be
brought up on our cluster.

\paragraph{TensorRT-LLM.}
TensorRT-LLM version 0.16 starts up a message passing layer as soon as it
is imported, and the message passing build on our nodes lacks the process
management support it expects. We tried four standard ways around this:
running it as a single process, launching one process with an isolated
process manager, disabling the automatic Python side initialization, and
launching it through the cluster scheduler with an older process manager,
which the scheduler blocked over step creation contention. All four
failed at initialization. This runtime's strongest documented advantages
are in any case realized on data center cards with a shared fast link,
which our cluster does not have.

\paragraph{SGLang.}
SGLang version 0.4.6 could not be installed against the combination on
our cluster: Python 3.12, CUDA 12.4 builds, our card generation, and the
transformers and compressed tensor versions the runtime pins. We made
eight attempts in both the host environment and a fresh isolated one, and
the dependency tree never resolved, in part because the recent prebuilt
kernels were published only for a newer card generation than ours. This
runtime's main advantage is a prompt cache that helps most when many
requests share a long common prefix, which our short synthetic prompts do
not exercise.

\paragraph{Scripts released.}
The benchmark scripts for both runtimes, covering 11 workloads each
with three run wrappers and full setup notes, are released with the code
so that the measurement can be repeated by readers who have a working
installation of either runtime on a compatible cluster.

\end{document}